\documentclass[aps,prl,twocolumn,showpacs,superscriptaddress,amsmath]{revtex4-1}

\pdfoutput=1

\usepackage{graphicx}
\usepackage{xspace}
\usepackage{hyperref} 
\usepackage[usenames]{color} 
\usepackage{ulem}
\usepackage{multirow}

\RequirePackage{lineno}

\begin{document}



\title{
Measurement of Muon Antineutrino Oscillations with an Accelerator-Produced Off-Axis Beam
}

\newcommand{\superk}           {Super-Kamiokande\xspace}       
\newcommand{\nue}                {$\nu_{e}$\xspace}
\newcommand{\nuebar}           {$\bar{\nu}_{e}$\xspace}
\newcommand{\numubar}           {$\bar{\nu}_{\mu}$\xspace}
\newcommand{\numu}             {$\nu_{\mu}$\xspace}
\newcommand{\nutau}             {$\nu_{\tau}$\xspace}
\newcommand{\nux}                {$\nu_{x}$\xspace}
\newcommand{\numunue}       {$\nu_{\mu} \rightarrow \nu_{e}$\xspace}
\newcommand{\numunux}       {$\nu_{\mu} \rightarrow \nu_{x}$\xspace}
\newcommand{\numunutau}    {$\nu_{\mu} \rightarrow \nu_{\tau}$\xspace}

\newcommand{\tonethree}       {$\theta_{13}$\xspace}
\newcommand{\tonetwo}         {$\theta_{12}$\xspace}
\newcommand{\ttwothree}       {$\theta_{23}$\xspace}
\newcommand{\ssttmue}          {$\sin^2 2 \theta_{{\mu}e}$\xspace}
\newcommand{\sstonethree}    {$\sin^2 2 \theta_{13}$\xspace}
\newcommand{\ssttwothree}    {$\sin^2 2 \theta_{23}$\xspace}

\newcommand{\msqonetwo}   {$\Delta m^2_{12}$\xspace}
\newcommand{\msqonethree} {$\Delta m^2_{13}$\xspace}
\newcommand{\msqtwothree} {$\Delta m^2_{23}$\xspace}
\newcommand{\absmsqtwothree} {$|\Delta m^2_{23}|$\xspace}
\newcommand{\msqmue}        {$\Delta m^2_{{\mu}e}$\xspace}
\newcommand{\msqmumu}     {$\Delta m^2_{\mu\mu}$\xspace}

\newcommand{\enu}               {$E_{\nu}$\xspace}
\newcommand{\pmu}              {$p_{\mu}$\xspace}
\newcommand{\amome}         {$E_{e}$\xspace}
\newcommand{\evis}              {$E_{vis}$\xspace}
\newcommand{\pizero}           {$\pi^{0}$\xspace}
\newcommand{\pizerogg}       {$\pi^{0}\to\gamma\gamma$\xspace}

\newcommand{\degree}        {$^\circ$\xspace}
\newcommand{\evsq}          {\mathrm{eV^2}}
\newcommand{\evmsq}          {\mathrm{eV^-2}}

\newcommand\brabar{\raisebox{-4.0pt}{\scalebox{.2}{
\textbf{(}}}\raisebox{-4.0pt}{{\_}}\raisebox{-4.0pt}{\scalebox{.2}{\textbf{)
}}}}


\newcommand{\INSTEE}{\affiliation{University of Bern, Albert Einstein Center for Fundamental Physics, Laboratory for High Energy Physics (LHEP), Bern, Switzerland}}
\newcommand{\INSTFE}{\affiliation{Boston University, Department of Physics, Boston, Massachusetts, U.S.A.}}
\newcommand{\INSTD}{\affiliation{University of British Columbia, Department of Physics and Astronomy, Vancouver, British Columbia, Canada}}
\newcommand{\INSTGA}{\affiliation{University of California, Irvine, Department of Physics and Astronomy, Irvine, California, U.S.A.}}
\newcommand{\INSTI}{\affiliation{IRFU, CEA Saclay, Gif-sur-Yvette, France}}
\newcommand{\INSTGB}{\affiliation{University of Colorado at Boulder, Department of Physics, Boulder, Colorado, U.S.A.}}
\newcommand{\INSTFG}{\affiliation{Colorado State University, Department of Physics, Fort Collins, Colorado, U.S.A.}}
\newcommand{\INSTFH}{\affiliation{Duke University, Department of Physics, Durham, North Carolina, U.S.A.}}
\newcommand{\INSTBA}{\affiliation{Ecole Polytechnique, IN2P3-CNRS, Laboratoire Leprince-Ringuet, Palaiseau, France }}
\newcommand{\INSTEF}{\affiliation{ETH Zurich, Institute for Particle Physics, Zurich, Switzerland}}
\newcommand{\INSTEG}{\affiliation{University of Geneva, Section de Physique, DPNC, Geneva, Switzerland}}
\newcommand{\INSTDG}{\affiliation{H. Niewodniczanski Institute of Nuclear Physics PAN, Cracow, Poland}}
\newcommand{\INSTCB}{\affiliation{High Energy Accelerator Research Organization (KEK), Tsukuba, Ibaraki, Japan}}
\newcommand{\INSTED}{\affiliation{Institut de Fisica d'Altes Energies (IFAE), The Barcelona Institute of Science and Technology, Campus UAB, Bellaterra (Barcelona) Spain}}
\newcommand{\INSTEC}{\affiliation{IFIC (CSIC \& University of Valencia), Valencia, Spain}}
\newcommand{\INSTEI}{\affiliation{Imperial College London, Department of Physics, London, United Kingdom}}
\newcommand{\INSTGF}{\affiliation{INFN Sezione di Bari and Universit\`a e Politecnico di Bari, Dipartimento Interuniversitario di Fisica, Bari, Italy}}
\newcommand{\INSTBE}{\affiliation{INFN Sezione di Napoli and Universit\`a di Napoli, Dipartimento di Fisica, Napoli, Italy}}
\newcommand{\INSTBF}{\affiliation{INFN Sezione di Padova and Universit\`a di Padova, Dipartimento di Fisica, Padova, Italy}}
\newcommand{\INSTBD}{\affiliation{INFN Sezione di Roma and Universit\`a di Roma ``La Sapienza'', Roma, Italy}}
\newcommand{\INSTEB}{\affiliation{Institute for Nuclear Research of the Russian Academy of Sciences, Moscow, Russia}}
\newcommand{\INSTHA}{\affiliation{Kavli Institute for the Physics and Mathematics of the Universe (WPI), The University of Tokyo Institutes for Advanced Study, University of Tokyo, Kashiwa, Chiba, Japan}}
\newcommand{\INSTCC}{\affiliation{Kobe University, Kobe, Japan}}
\newcommand{\INSTCD}{\affiliation{Kyoto University, Department of Physics, Kyoto, Japan}}
\newcommand{\INSTEJ}{\affiliation{Lancaster University, Physics Department, Lancaster, United Kingdom}}
\newcommand{\INSTFC}{\affiliation{University of Liverpool, Department of Physics, Liverpool, United Kingdom}}
\newcommand{\INSTFI}{\affiliation{Louisiana State University, Department of Physics and Astronomy, Baton Rouge, Louisiana, U.S.A.}}
\newcommand{\INSTJ}{\affiliation{Universit\'e de Lyon, Universit\'e Claude Bernard Lyon 1, IPN Lyon (IN2P3), Villeurbanne, France}}
\newcommand{\INSTHB}{\affiliation{Michigan State University, Department of Physics and Astronomy,  East Lansing, Michigan, U.S.A.}}
\newcommand{\INSTCE}{\affiliation{Miyagi University of Education, Department of Physics, Sendai, Japan}}
\newcommand{\INSTDF}{\affiliation{National Centre for Nuclear Research, Warsaw, Poland}}
\newcommand{\INSTFJ}{\affiliation{State University of New York at Stony Brook, Department of Physics and Astronomy, Stony Brook, New York, U.S.A.}}
\newcommand{\INSTGJ}{\affiliation{Okayama University, Department of Physics, Okayama, Japan}}
\newcommand{\INSTCF}{\affiliation{Osaka City University, Department of Physics, Osaka, Japan}}
\newcommand{\INSTGG}{\affiliation{Oxford University, Department of Physics, Oxford, United Kingdom}}
\newcommand{\INSTBB}{\affiliation{UPMC, Universit\'e Paris Diderot, CNRS/IN2P3, Laboratoire de Physique Nucl\'eaire et de Hautes Energies (LPNHE), Paris, France}}
\newcommand{\INSTGC}{\affiliation{University of Pittsburgh, Department of Physics and Astronomy, Pittsburgh, Pennsylvania, U.S.A.}}
\newcommand{\INSTFA}{\affiliation{Queen Mary University of London, School of Physics and Astronomy, London, United Kingdom}}
\newcommand{\INSTE}{\affiliation{University of Regina, Department of Physics, Regina, Saskatchewan, Canada}}
\newcommand{\INSTGD}{\affiliation{University of Rochester, Department of Physics and Astronomy, Rochester, New York, U.S.A.}}
\newcommand{\INSTHC}{\affiliation{Royal Holloway University of London, Department of Physics, Egham, Surrey, United Kingdom}}
\newcommand{\INSTBC}{\affiliation{RWTH Aachen University, III. Physikalisches Institut, Aachen, Germany}}
\newcommand{\INSTFB}{\affiliation{University of Sheffield, Department of Physics and Astronomy, Sheffield, United Kingdom}}
\newcommand{\INSTDI}{\affiliation{University of Silesia, Institute of Physics, Katowice, Poland}}
\newcommand{\INSTEH}{\affiliation{STFC, Rutherford Appleton Laboratory, Harwell Oxford,  and  Daresbury Laboratory, Warrington, United Kingdom}}
\newcommand{\INSTCH}{\affiliation{University of Tokyo, Department of Physics, Tokyo, Japan}}
\newcommand{\INSTBJ}{\affiliation{University of Tokyo, Institute for Cosmic Ray Research, Kamioka Observatory, Kamioka, Japan}}
\newcommand{\INSTCG}{\affiliation{University of Tokyo, Institute for Cosmic Ray Research, Research Center for Cosmic Neutrinos, Kashiwa, Japan}}
\newcommand{\INSTGI}{\affiliation{Tokyo Metropolitan University, Department of Physics, Tokyo, Japan}}
\newcommand{\INSTF}{\affiliation{University of Toronto, Department of Physics, Toronto, Ontario, Canada}}
\newcommand{\INSTB}{\affiliation{TRIUMF, Vancouver, British Columbia, Canada}}
\newcommand{\INSTG}{\affiliation{University of Victoria, Department of Physics and Astronomy, Victoria, British Columbia, Canada}}
\newcommand{\INSTDJ}{\affiliation{University of Warsaw, Faculty of Physics, Warsaw, Poland}}
\newcommand{\INSTDH}{\affiliation{Warsaw University of Technology, Institute of Radioelectronics, Warsaw, Poland}}
\newcommand{\INSTFD}{\affiliation{University of Warwick, Department of Physics, Coventry, United Kingdom}}
\newcommand{\INSTGE}{\affiliation{University of Washington, Department of Physics, Seattle, Washington, U.S.A.}}
\newcommand{\INSTGH}{\affiliation{University of Winnipeg, Department of Physics, Winnipeg, Manitoba, Canada}}
\newcommand{\INSTEA}{\affiliation{Wroclaw University, Faculty of Physics and Astronomy, Wroclaw, Poland}}
\newcommand{\INSTH}{\affiliation{York University, Department of Physics and Astronomy, Toronto, Ontario, Canada}}

\INSTEE
\INSTFE
\INSTD
\INSTGA
\INSTI
\INSTGB
\INSTFG
\INSTFH
\INSTBA
\INSTEF
\INSTEG
\INSTDG
\INSTCB
\INSTED
\INSTEC
\INSTEI
\INSTGF
\INSTBE
\INSTBF
\INSTBD
\INSTEB
\INSTHA
\INSTCC
\INSTCD
\INSTEJ
\INSTFC
\INSTFI
\INSTJ
\INSTHB
\INSTCE
\INSTDF
\INSTFJ
\INSTGJ
\INSTCF
\INSTGG
\INSTBB
\INSTGC
\INSTFA
\INSTE
\INSTGD
\INSTHC
\INSTBC
\INSTFB
\INSTDI
\INSTEH
\INSTCH
\INSTBJ
\INSTCG
\INSTGI
\INSTF
\INSTB
\INSTG
\INSTDJ
\INSTDH
\INSTFD
\INSTGE
\INSTGH
\INSTEA
\INSTH

\author{K.\,Abe}\INSTBJ
\author{C.\,Andreopoulos}\INSTEH\INSTFC
\author{M.\,Antonova}\INSTEB
\author{S.\,Aoki}\INSTCC
\author{A.\,Ariga}\INSTEE
\author{S.\,Assylbekov}\INSTFG
\author{D.\,Autiero}\INSTJ
\author{M.\,Barbi}\INSTE
\author{G.J.\,Barker}\INSTFD
\author{G.\,Barr}\INSTGG
% ...

\collaboration{The T2K Collaboration}\noaffiliation

\date{\today}

\begin{abstract}
T2K reports its first measurements of the parameters governing the disappearance of $\overline{\nu}_\mu$ in an off-axis beam due to flavor change induced by neutrino oscillations. The quasimonochromatic $\overline{\nu}_\mu$ beam, produced with a peak energy of 0.6 GeV at J-PARC, is observed at the far detector Super-Kamiokande, 295~km away, where the $\overline{\nu}_\mu$ survival probability is expected to be minimal. Using a dataset corresponding to $4.01 \times 10^{20}$ protons on target, $34$ fully contained $\mu$-like events were observed. The best-fit oscillation parameters are $\sin^2 (\overline{\theta}_{23}) = 0.45$ and $|\Delta\overline{m}^2_{32}| = 2.51 \times 10^{-3}$~eV$^2$ with 68\% confidence intervals of 0.38 - 0.64 and 2.26 - 2.80 $\times 10^{-3}$~eV$^2$ respectively. These results are in agreement with existing antineutrino parameter measurements and also with the $\nu_\mu$ disappearance parameters measured by T2K.
\end{abstract}

\maketitle

{\it Introduction.---}In the three-flavor framework, neutrino oscillation can be described by the unitary Pontecorvo- Maki-Nakagawa-Sakata matrix, which is parameterized by three angles $\theta_{12}, \theta_{23}, \theta_{13}$ and a 
$CP$-violating phase $\delta_{CP}$~\cite{Maki:1962mu,Pontecorvo:1967fh,PhysRevD.86.010001neutrino}.
Given a neutrino propagation distance, $L$~(km), and energy, $E_{\nu}$~(GeV), such that $L/E_{\nu} \sim O(1000)$, the survival probability for a muon neutrino propagating through vacuum can be approximated by:
\begin{equation}
\begin{split}
    P(\nu_{\mu} & \rightarrow \nu_{\mu}) \simeq 1 - 4\cos^{2}(\theta_{13})\sin^{2}(\theta_{23}) \times \\
                            & [1-\cos^{2}(\theta_{13})\sin^{2}(\theta_{23})] \sin^{2}(\frac{1.267\Delta m^{2}_{32}L}{E_{\nu}}),
\end{split}
\label{equ:neutrino_oscillation_prob}
\end{equation}
where $\Delta m^{2}_{32}~\textrm{(eV}^{2}\textrm{)}$ is the neutrino mass squared splitting, defined as $m^{2}_{3} - m^{2}_{2}$. 
Equation~(\ref{equ:neutrino_oscillation_prob}) shows that measuring the disappearance probability as a function of $L/E_{\nu}$ leads to a measurement of the oscillation parameters.
In this model of neutrino oscillation, the disappearance probability in vacuum is identical for neutrinos and antineutrinos. 
The disappearance probabilities in matter 
can differ by as much as 0.1\% for the T2K baseline and neutrino flux, but  
our dataset is not sensitive to this small effect.
Observing a significant difference between the disappearance probabilities of neutrinos and antineutrinos would, therefore, be evidence for 
new physics~\cite{PhysRevD.86.010001neutrino}.
Results from the MINOS~\cite{Adamson:2012rm} and Super-Kamiokande (SK)
Collaborations~\cite{Abe:2011ph} 
indicate no significant difference between muon antineutrino oscillations and muon neutrino oscillations.

In this Letter we present the first measurement of $P(\overline{\nu}_{\mu} \rightarrow \overline{\nu}_{\mu})$ by the T2K Collaboration. 
This analysis allows 
the dominant antineutrino oscillation parameters for $\overline{\nu}_\mu$ 
disappearance
to vary independently from
those describing neutrino oscillations, {\it i.e.} $\theta_{23} \neq \overline{\theta}_{23}$ and $\Delta m^{2}_{32} \neq \Delta \overline{m}^{2}_{32}$, where the barred parameters refer to antineutrino oscillations.
$\overline{\theta}_{13}$, $\overline{\theta}_{12}$ and $\Delta \overline{m}^{2}_{21}$ are assumed to be identical to their matter counterparts since our dataset cannot constrain them.
This ensures that the expected background at the far detector is consistent with the current knowledge of neutrino oscillations, while allowing us to use the T2K antineutrino-mode data to measure $\overline{\theta}_{23}$ and $\Delta \overline{m}^{2}_{32}$.

{\it T2K experiment.---}The T2K experiment~\cite{Abe:2011ks} is composed of a neutrino beam line, a suite of near detectors, and the far detector, Super-Kamiokande.
Both the far detector and one of the near detectors are placed 2.5$^{\circ}$ off the neutrino beam axis and so observe a narrow-band beam~\cite{beavis:bnl}.
This ``off-axis'' method reduces backgrounds from higher-energy neutrinos, 
producing a neutrino flux that peaks around 0.6~GeV,
the energy
at which
the first minimum in the $\overline{\nu}_{\mu}$ survival probability is expected to occur at the T2K baseline.

The J-PARC main ring provides a 30-GeV proton beam which impinges upon a graphite target, producing pions and kaons.
The target is held inside the first of three magnetic horns which focus charged particles into a 96-m-long decay volume, where they decay and produce neutrinos.
The polarity of the horn current determines whether positive or negative mesons are focused, which in turn determines whether the neutrino beam is largely composed of muon neutrinos or muon antineutrinos.
The decay volume ends in a beam dump followed by the muon monitor,
which measures the neutrino beam direction on a bunch-by-bunch basis using muons from the meson decays.

The near detector complex~\cite{Abe:2011ks}  
consists of the on-axis Interactive Neutrino GRID detector (INGRID)~\cite{Abe2012} 
and the off-axis detector (ND280), both 280~m downstream of the proton-beam target.
INGRID is a 7+7 array of iron-scintillator detectors, arranged in a ``cross'' configuration at the beam center. 
INGRID provides high-statistics monitoring of the neutrino beam intensity, direction, profile, and stability and has shown that the neutrino beam direction is controlled to 0.4~mrad.
ND280 consists of a number of subdetectors installed inside the refurbished UA1/NOMAD magnet, which provides a 0.2~T field. 
The near detector analysis 
described here
uses the tracker region of ND280, which consists of three time projection chambers (TPC1, 2, 3)~\cite{Abgrall:2010hi}
interleaved with two fine-grained detectors (FGD1, 2)~\cite{Amaudruz:2012pe}.
The FGDs are the neutrino target and track charged particles coming from the interaction vertex,
while the TPCs perform 3D tracking and determine the charge, momentum, and energy loss of each charged particle traversing them.
The observed energy loss is used for particle identification which, when combined with particle charge information,
allows a precise separation and measurement of the $\overline{\nu}_\mu$ (right-sign) and $\nu_\mu$ (wrong-sign) interactions in the antineutrino-mode beam.

The far detector is a 50-kt 
(22.5-kt fiducial mass) 
water Cherenkov detector~\cite{Fukuda2003418,Abe20140211} 
where
the volume is divided into an outer detector (OD) with 1885 outward-facing
20-cm-diameter photomultiplier tubes and an inner detector (ID) with 11\,129 
inward-facing 50-cm-diameter photomultiplier tubes.  The events arriving at SK from the J-PARC beam spill
are synchronized with a global positioning system with $<150$~ns precision.

The results presented here are 
based on data taken in three periods: 
two where the beam operated in antineutrino mode, (1) June 2014 and (2) November 2014 -- June 2015,
and one in neutrino mode, (3) November 2010 -- May 2013.
The oscillation analysis uses periods (1) and (2), while 
the near-detector analysis uses data from periods (1) and (3).
This corresponds to 
an exposure of $4.01 \times 10^{20}$ protons on target (POT) in antineutrino mode for the oscillation analysis, and
an exposure of $0.43 \times 10^{20}$ POT in antineutrino mode plus $5.82 \times 10^{20}$ POT in neutrino mode for the near-detector analysis.

{\it Analysis strategy.---}This analysis resembles that of Ref.~\cite{Abe:2015awa}, 
fitting samples of charged-current (CC) interactions at ND280 to produce 
a tuned prediction of the unoscillated antineutrino spectrum at the far detector,
including its associated uncertainty.
This analysis differs from Ref.~\cite{Abe:2015awa} in that 
both $\nu_{\mu}$ and $\overline{\nu}_{\mu}$ samples at ND280 
are fitted.  This ensures that the neutrino interaction model is consistent between both neutrino- 
and antineutrino-beam-mode data sets and provides a constraint on 
both the right-sign signal and the wrong-sign background 
in the antineutrino-mode beam.

\begin{figure}
\centering
\includegraphics[width=8cm]{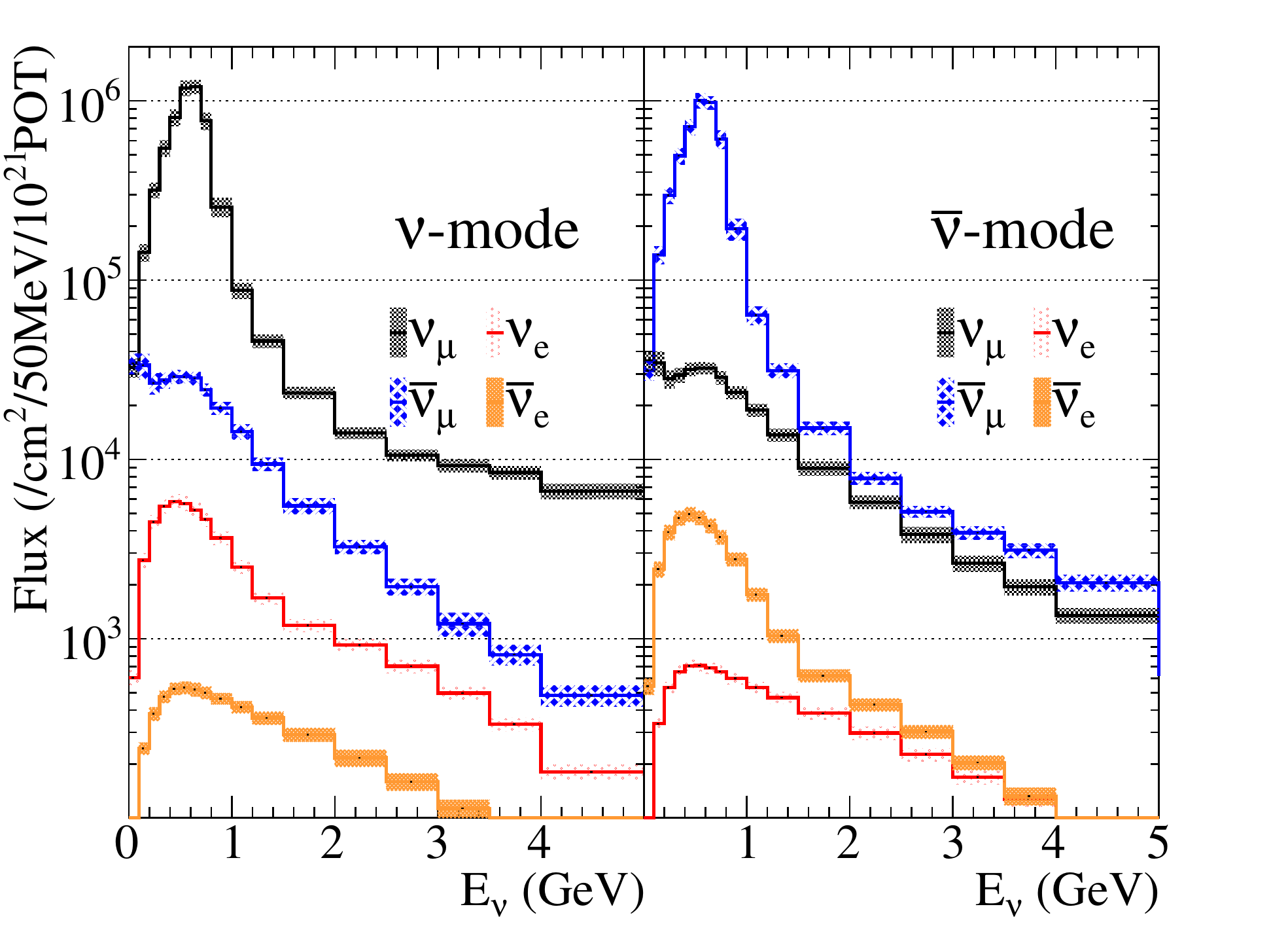}
\caption{The nominal unoscillated neutrino flux prediction at SK for each neutrino type in the neutrino-mode beam (left) and antineutrino-mode beam (right).  The shaded boxes indicate the total systematic uncertainty on each energy bin.}
\label{fig:flux}
\end{figure}

{\it Flux simulation.---}The nominal neutrino flux at ND280 and SK (without oscillation) is predicted by simulating the
secondary beamline~\cite{Abe:2012av} using FLUKA2011~\cite{cite:FLUKA1,FLUKA2011}
and GEANT3 with GCALOR~\cite{GEANT3, cite:GCALOR}.
The simulated hadronic interactions are tuned to external hadron-production data.
The unoscillated neutrino flux prediction at SK is shown in Fig.~\ref{fig:flux} for each neutrino type
and for both neutrino- and antineutrino-mode beams.
At the peak energy of the T2K beam, the $\nu_\mu$ flux in the neutrino-mode 
beam is 20\% higher than the $\overline{\nu}_\mu$ flux in the 
antineutrino-mode beam,
due to the larger production cross section for $\pi^+$ compared to $\pi^-$ 
in proton-carbon interactions.
The ratio of the wrong-sign component ($\nu_{\mu}$ in the
$\overline{\nu}_\mu$ beam), 
mainly coming from forward-going high-energy pions,
 to the right-sign component ($\overline\nu_{\mu}$) at the peak energy is 3\%.
The largest sources of neutrino flux uncertainty are from beam-line and hadron-production modeling uncertainties, which are common to ND280 and SK.
The new NA61/SHINE 2009 thin-target data~\cite{Abgrall:2015xoa} are included in the hadron-production tuning for this analysis, reducing the total flux uncertainty from between 12\%-15\% to 10\% around 0.6~GeV.

{\it Neutrino interaction simulation.---}Neutrino interactions are modeled with the NEUT Monte Carlo event generator~\cite{Hayato:2009zzz,Hayato:2009zzzz,Nieves:2011pp,Nieves:2004wx,cite:Wilkinson:2015}. 
The generator uses the same model with common parameters to describe both $\nu$ and $\bar\nu$ interactions.
In the case of CC quasi-elastic reactions (CCQE: $\nu_\mu + n\rightarrow \mu^- +p$ or $\overline{\nu}_\mu + p\rightarrow \mu^+ + n$)
neutrino and antineutrino cross sections differ by the sign of the vector-axial interference term~\cite{LlewellynSmith:1971zm,Jacob}.
At a neutrino energy of 0.6~GeV, this makes the neutrino-oxygen CCQE cross section a factor of $\sim4$ larger than that of antineutrinos.

To set the initial values and uncertainties of some parameters, such as the
CCQE axial mass and the normalization of the multinucleon contribution, 
results from the MiniBooNE and MINER$\nu$A experiments~\cite{AguilarArevalo:2010zc, AguilarArevalo:2013hm, Fiorentini:2013ezn, Fields:2013zhk}
on CH$_2$ and CH targets are used.
These parameters are then tuned by the near-detector fit.

{\it Near detector fit.---}
The seven samples used in the near-detector fit
are summarized in Table~\ref{table:ndres}. 
\begin{table}[!tbp]
  \centering
  \caption{Data and MC-predicted event rates for the different ND280 samples before and after the fit. Errors indicate systematic uncertainties only.}
  \label{table:ndres}
  \begin{tabular*}{\columnwidth}{l@{\extracolsep{\fill}}c@{\extracolsep{\fill}}c@{\extracolsep{\fill}}c}
   \hline
   \hline
  Sample               & Data &  Prefit &  Postfit \\
   \hline
   \multicolumn{4}{c}{$\nu$ beam mode}\\
$\nu_\mu$ CC 0$\pi$ & 17\,362 & 15\,625 $\pm$ 1663 & 17\,248 $\pm$ 133 \\
$\nu_\mu$ CC 1$\pi^{+}$ & 3988  & 4748  $\pm$ 686  & 4190  $\pm$ 60 \\
$\nu_\mu$ CC other  & 4219  & 3772  $\pm$ 431  & 4079  $\pm$ 62 \\
   \multicolumn{4}{c}{$\bar\nu$ beam mode}\\
$\overline{\nu}_\mu$ CC 1 track  & 435 & 387 $\pm$ 41 & 438 $\pm$ 13 \\
$\overline{\nu}_\mu$ CC N tracks & 136 & 128 $\pm$ 17 & 129 $\pm$ 5 \\
$\nu_\mu$ CC 1 track             & 131 & 141 $\pm$ 15 & 147 $\pm$ 6 \\
$\nu_\mu$ CC N tracks            & 145 & 147 $\pm$ 17 & 144 $\pm$ 6 \\
   \hline
   \hline
  \end{tabular*}
  \end{table}
Muon-neutrino-induced CC interactions 
in the neutrino beam mode
are found by requiring that the highest-momentum, negative-curvature track in an event starts within the
upstream FGD (FGD1) fiducial volume (FV) and has an energy deposit in TPC2 consistent with a muon.
Events with a TPC track that starts upstream of the start point of the muon candidate are rejected 
and the remaining $\nu_\mu$ CC candidates are divided into three subsamples according to the number of associated pions: 
$\nu_\mu$ CC 0$\pi$, $\nu_\mu$ CC 1$\pi^+$, and $\nu_\mu$ CC other,
which are dominated by CCQE, CC resonant pion production, and deep inelastic scattering interactions, respectively~\cite{Abe:2015awa}.
For the antineutrino-beam-mode samples,
the selection of $\overline{\nu}_\mu$ ($\nu_\mu$) CC interactions is similar to that used in the neutrino beam mode, except 
the positive (negative) track must be the highest-momentum track in the event.
The selected $\overline{\nu}_\mu$ ($\nu_\mu$) CC candidate events are divided into two subsamples rather than three, due to the small amount of antineutrino-mode data used in this analysis.
These are defined by the number of reconstructed tracks crossing TPC2: 
$\overline{\nu}_\mu$ ($\nu_\mu$) 
CC 1 track, dominated by CCQE interactions; and 
$\overline{\nu}_\mu$ ($\nu_\mu$) 
CC N tracks (N $>$ 1), 
a mixture of resonant production and deep inelastic scattering.

\begin{figure}[!tbp]
\centering
\leftline{\includegraphics[trim=0 2.85cm 0 0,clip,width=0.5\linewidth]{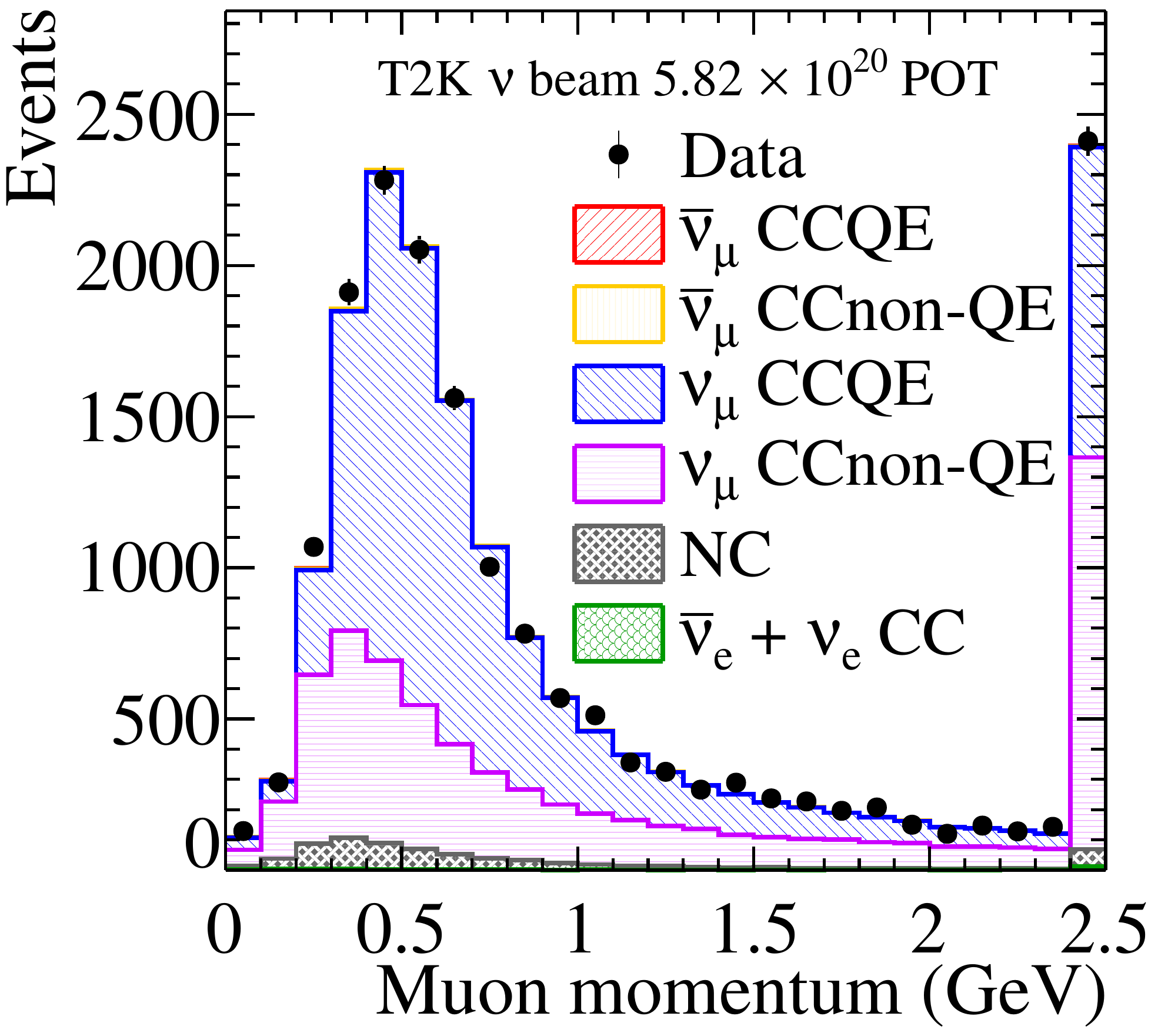}
\includegraphics[trim=0 2.85cm 0 0,clip,width=0.5\linewidth]{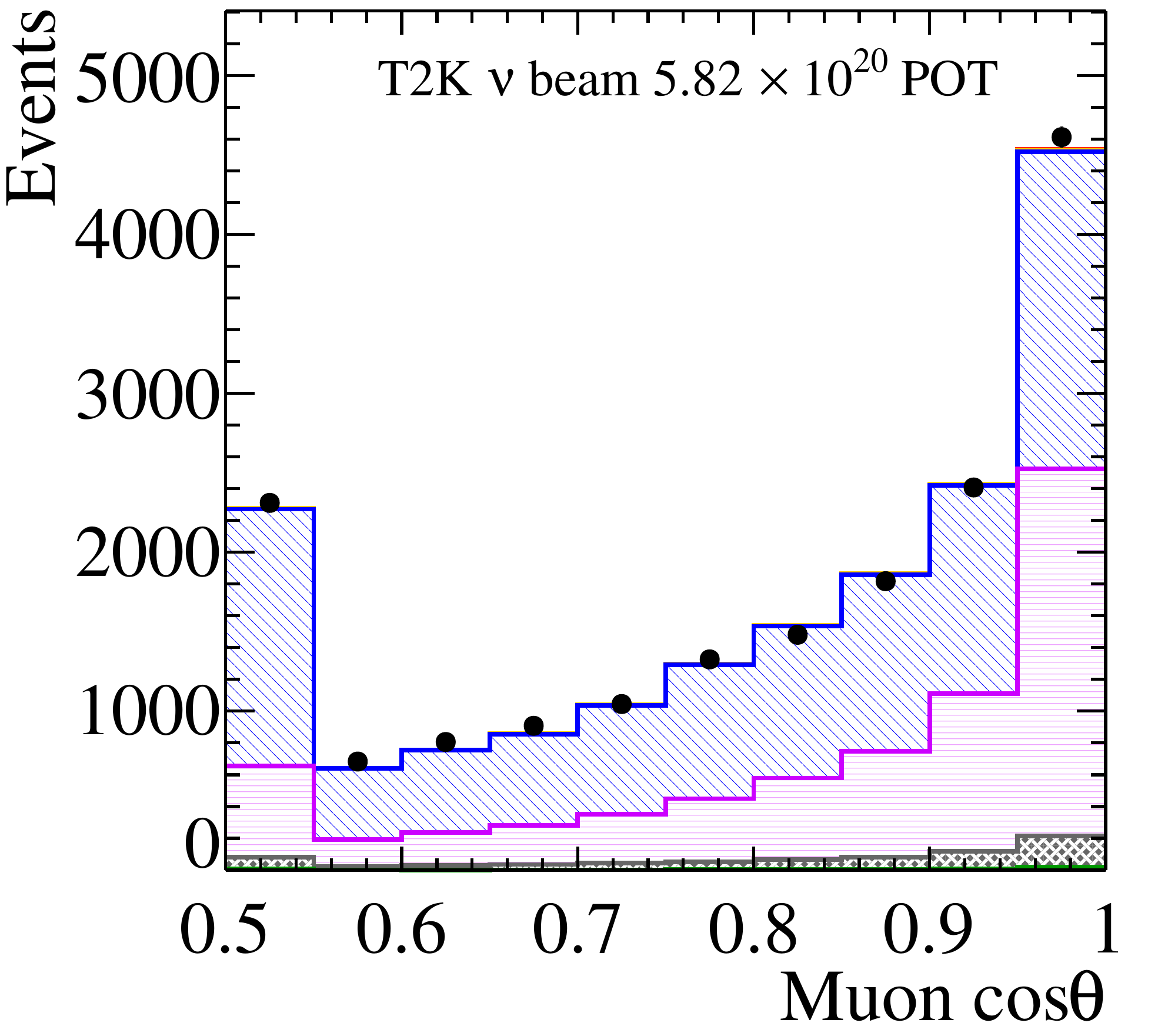}}
\vspace{-0.03cm}
\leftline{\includegraphics[trim=0 2.85cm 0 0.20cm,clip,width=0.5\linewidth]{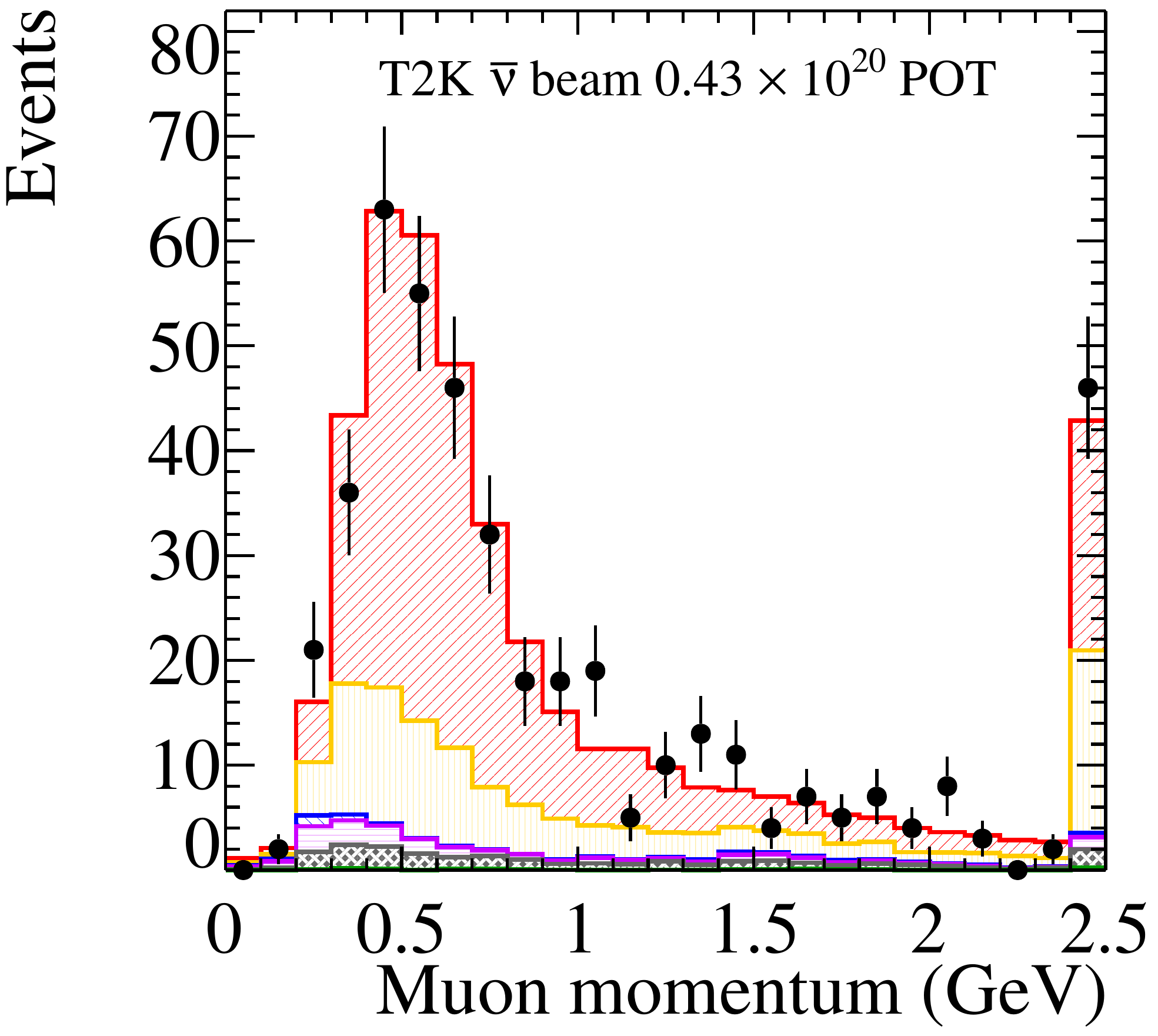}
\includegraphics[trim=0 2.85cm 0 0.20cm,clip,width=0.5\linewidth]{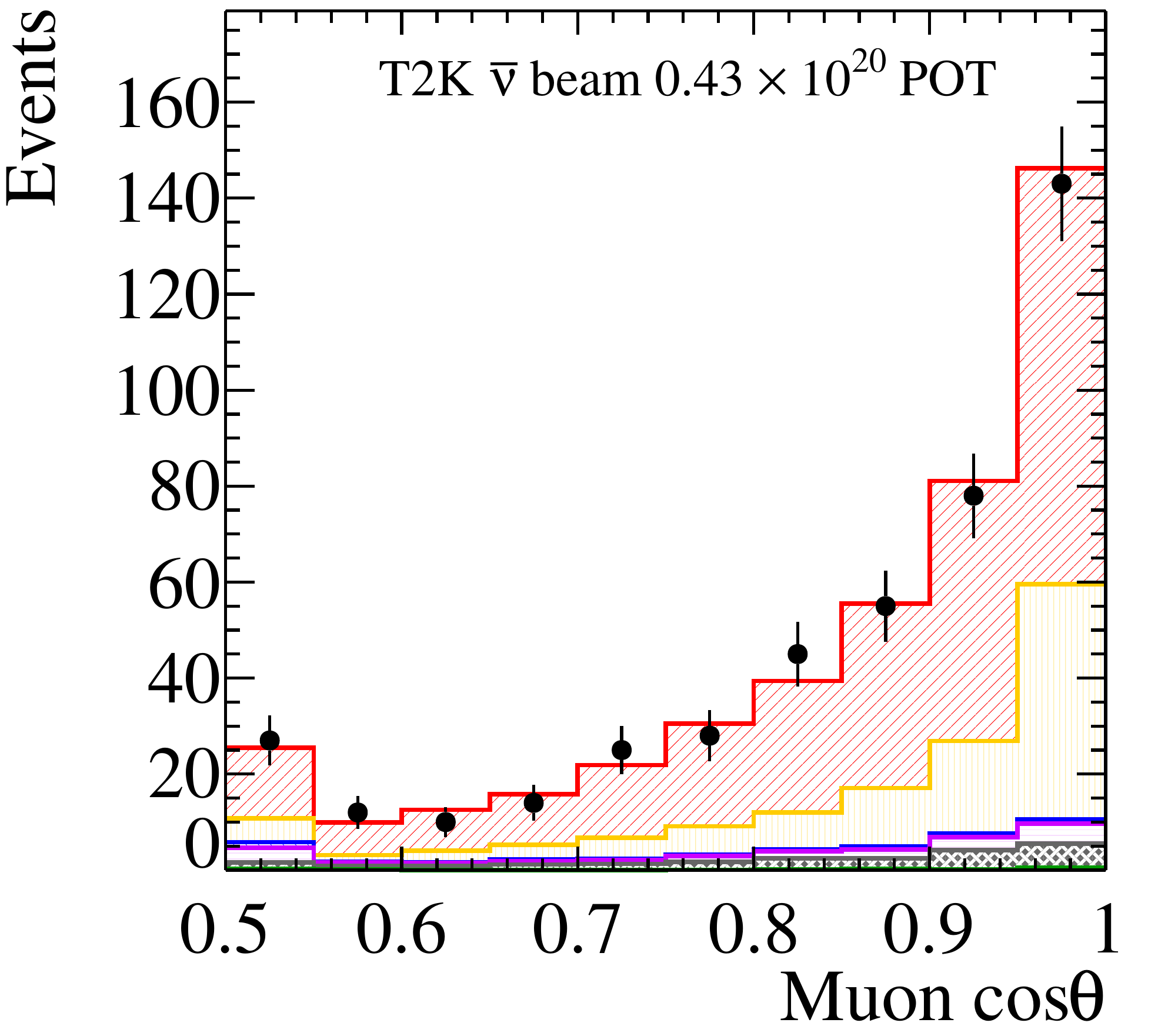}}
\vspace{-0.03cm}
\leftline{\includegraphics[trim=0 0 0 0.20cm,clip,width=0.5\linewidth]{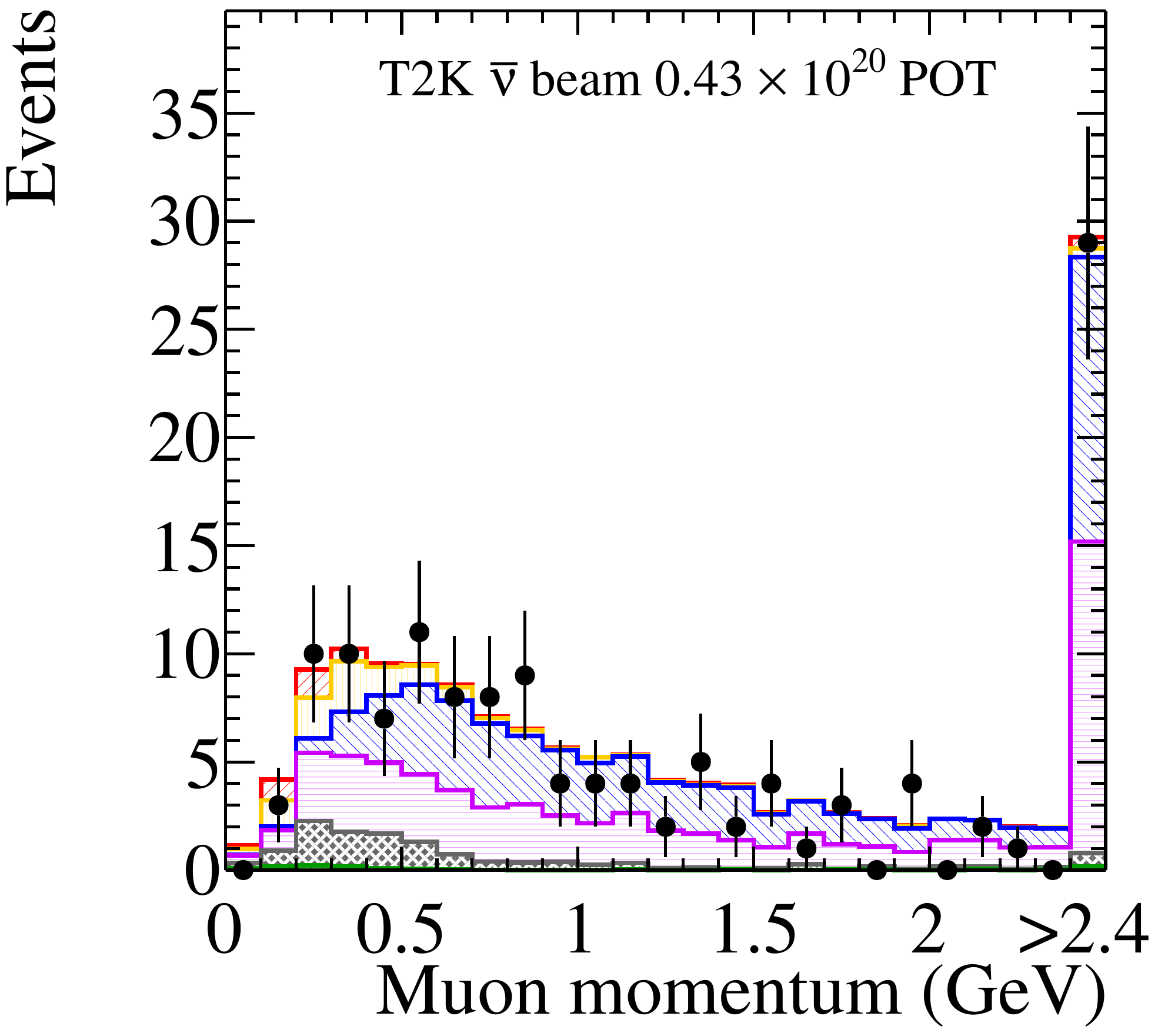}
\includegraphics[trim=0 0 0 0.20cm,clip,width=0.5\linewidth]{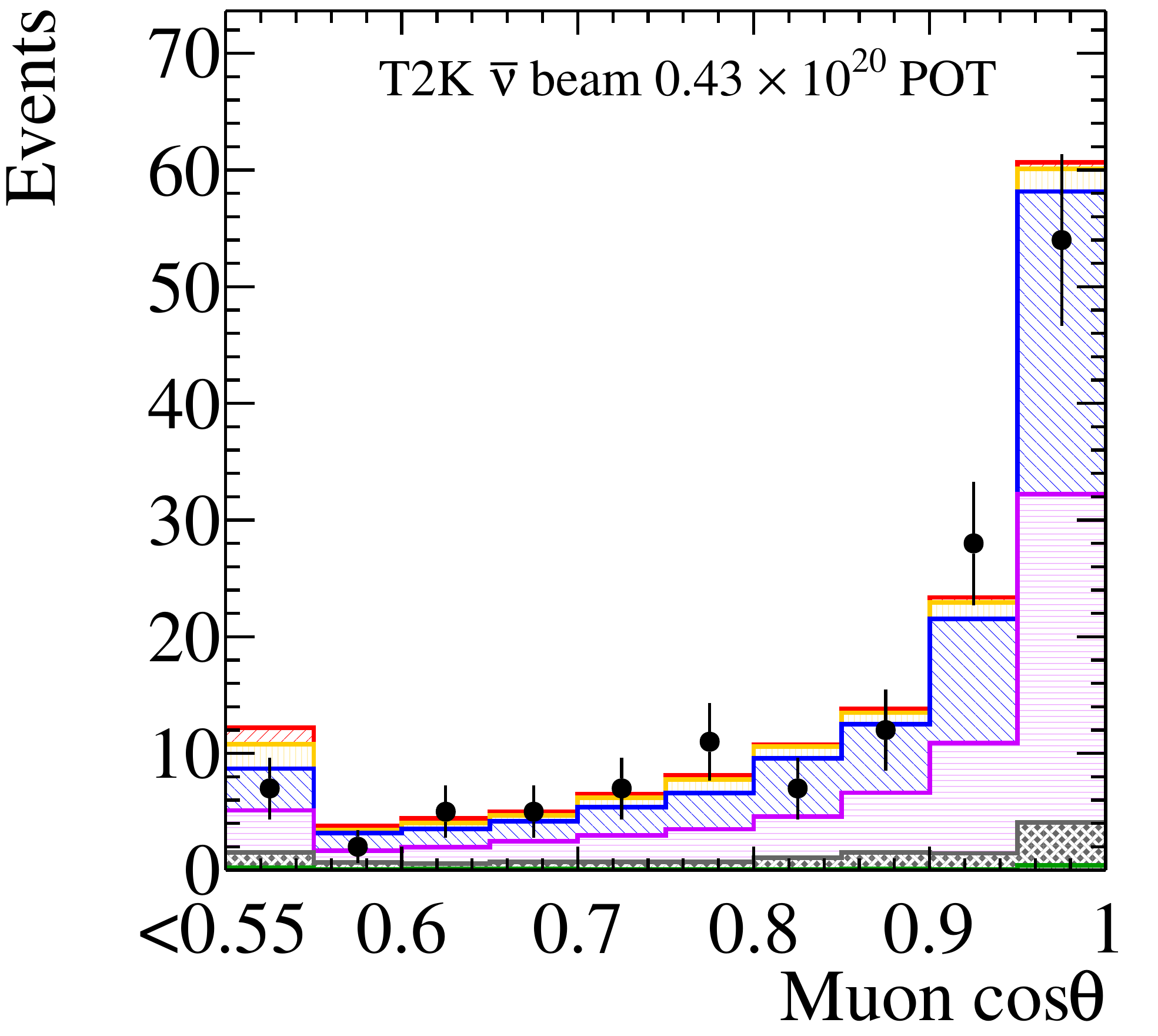}}
\caption{The momentum (left) and angular (right) distributions of the muon candidates at ND280 from the $\nu_\mu$ CC 0$\pi$ (top), the $\overline{\nu}_\mu$ CC 1 track (center) and the $\nu_\mu$ CC 1 track (bottom) samples. The data are superimposed on the post-fit MC prediction, separated by interaction mode.}
\label{fig:ndnures}
\end{figure}

The fit uses a binned likelihood, with the samples binned according to the muon momentum
and angle ($\theta$) relative to the 
central axis of the detector, roughly 1.7\degree away from the
incident (anti)neutrino direction.  The TPCs calculate the muon momentum from the curvature of the lepton in the ND280 magnetic field, with a resolution of 6\% at 1~GeV/$c$~\cite{Abgrall:2010hi}.
Figure~\ref{fig:ndnures} shows the 1D projections of these distributions for the $\nu_\mu$ CC 0$\pi$, the $\overline{\nu}_\mu$ CC 1 track, and the $\nu_\mu$ CC 1 track
samples for both data and the post-fit MC model.
The $p$-value of the data fit likelihood ratio was found to be 0.05, and the agreement between the ND280 data and the MC model was judged to be acceptable.
The fit gives estimates for 25 antineutrino beam flux parameters at SK, 12 cross-section parameters (including 4 specific to oxygen), and their covariance.
There are also additional parameters to control pion final state interactions (FSI) and reinteractions within the detector, which are independent for ND280 and SK.

To decouple the properties of the carbon target at ND280 from those of the oxygen target at SK, separate Fermi momentum, binding energy, multinucleon event normalization, and CC coherent pion-production normalization parameters are introduced for interactions on oxygen.
Since oxygen comprises only 3.6\% by mass of the FGD1 target, this near-detector analysis is insensitive to these parameters.
A conservative (100\% uncertainty) ansatz is adopted for the normalization of multinucleon ejection oxygen events, giving a 9.5\% uncertainty on the number of events at SK.
For the parameters that ND280 can constrain, the fit reduces their effect on
the uncertainty on the expected number of events at SK from $9.2\%$ to $3.4\%$.

{\it Far detector selection.---}At the far detector,
fully contained fiducial volume (FCFV) events are selected by requiring 
no hit clusters in the OD, 
that the reconstructed interaction vertex is more than 2~m away from 
the ID wall,
and that the visible energy in the event is larger than 30~MeV.
The last criterion requires that the amount of Cherenkov light is
more than that of a 30~MeV electromagnetic shower. 

\begin{figure}
\centering
\includegraphics[width=8cm]{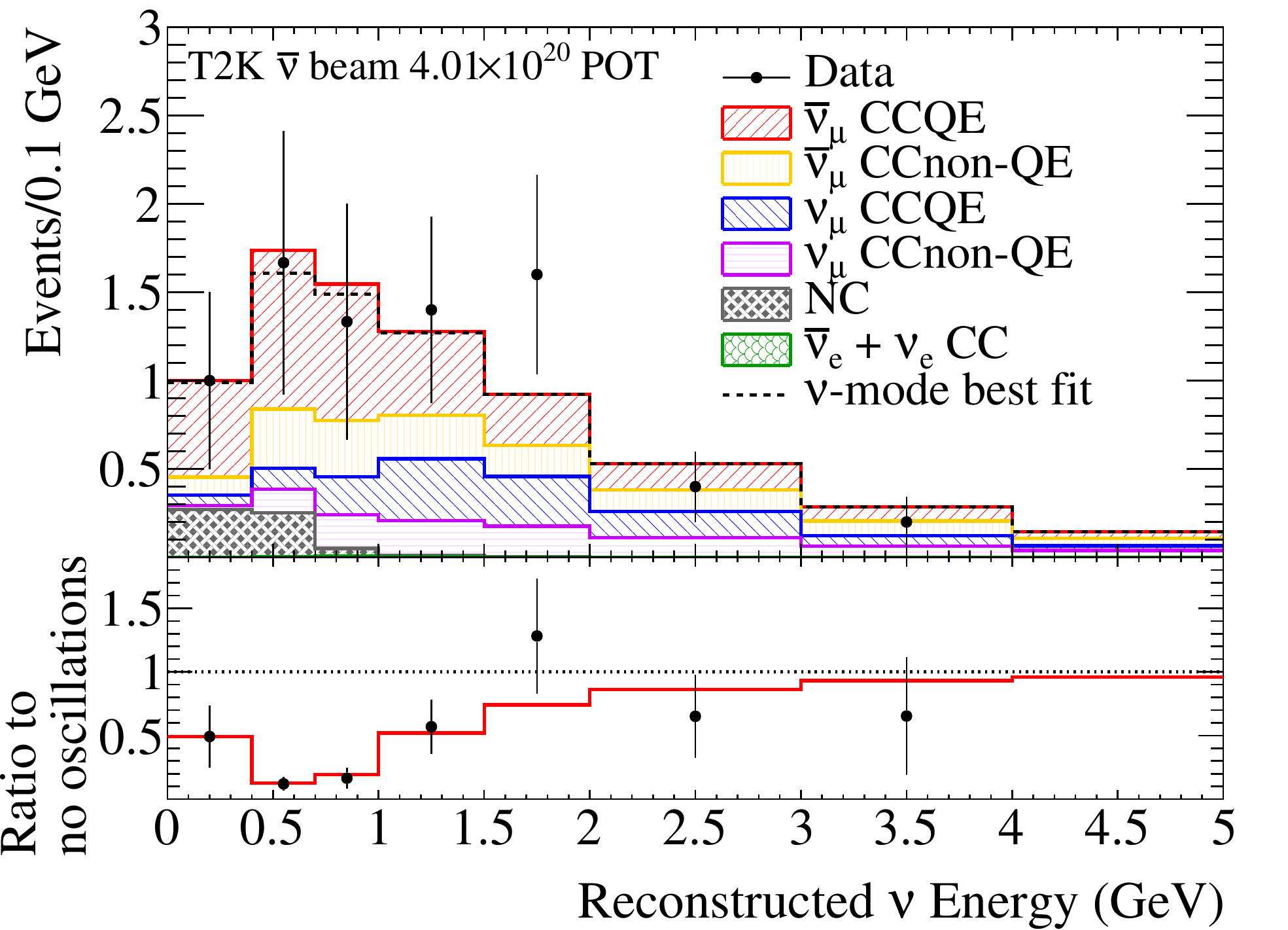}
\caption{Top: The reconstructed energy distribution of the 34 far-detector $\overline{\nu}_\mu$ candidates 
and the best-fit prediction, separated by interaction mode.
This is compared to the predicted spectrum assuming the antineutrino oscillation parameters are identical to the neutrino parameters measured by T2K~\cite{Abe:2015awa}.
Bottom: The observed data and $\overline{\nu}_\mu$-mode
best-fit prediction as a ratio to the unoscillated prediction.}
\label{fig:erec_ratio}
\end{figure}

To enhance the  
$\overline{\nu}_\mu$ 
CCQE purity of the sample, selected events must have a single, $\mu$-like Cherenkov ring,
no more than one decay electron,
and a muon momentum greater than 0.2~GeV~\cite{Abe:2015awa}. 
The number of data and MC events passing each selection criterion are shown 
in Table~\ref{table:event_reduction} and the reconstructed energy spectrum of the 34 selected events is plotted in Fig.~\ref{fig:erec_ratio}.
The reconstructed neutrino energy is calculated using the muon momentum 
and production angle, under the assumption that a CCQE interaction occurred on a nucleon at rest.
The selection efficiency for $\overline{\nu}_\mu$ CCQE is estimated to be 77\% while backgrounds from neutral-current (NC), $\nu_e$, and $\overline{\nu}_e$ interactions are reduced by a factor of 50.
The systematic uncertainties in the detector response 
are evaluated using atmospheric 
neutrinos, cosmic-ray muons, and their decay electrons~\cite{Abe:2015awa}.

\begin{table}
  \caption{ The number of events observed at the far detector in the antineutrino-beam-mode data 
  after applying each selection cut.
  MC expectation is calculated assuming oscillations with 
  $\sin^2(\theta_{23}) = \sin^2(\overline{\theta}_{23}) = 0.5$,
  $|\Delta m^2_{32}|=|\Delta \overline{m}^2_{32}| = 2.4\times10^{-3}~\textrm{eV}^2$, 
  and $\sin^2(\theta_{13}) = \sin^2(\overline{\theta}_{13}) = 0.0257$.
  The ``$\overline{\nu}_{e}$+\nue+NC'' column includes the NC interactions of all the (anti)neutrino flavors.
  Efficiency numbers are calculated with respect to the number of MC events
  generated in the fiducial volume (FV interaction).}
  \label{table:event_reduction}
  \begin{tabular}{lccccccc}
   \hline
   \hline
                                & \multirow{2}{*}{Data\;\;} & Total & \multicolumn{2}{c}{CCQE} & \multicolumn{2}{c}{CCnonQE} & \;\;$\overline{\nu}_{e}$+\nue \\
                                &      & MC     &  $\overline{\nu}_{\mu}$ & ${\nu}_{\mu}$ & \;\;$\overline{\nu}_{\mu}$ & \numu & +NC \\
   \hline
   FV interaction               &  $\cdot\cdot\cdot$ &  186.7	&  17.8   & 11.4 &  20.0 & 36.5  & 101    \\
   FCFV							&  90  &  99.7 	&  14.4   & 8.6  &  15.1 & 26.6  & 35.1    \\
   Single ring          		&  50  &  52.2 	&  14.0   & 7.7  &  8.1 & 8.7    & 13.8    \\
   $\mu$-like           		&  40  &  39.4 	&  13.8   & 7.6  &  7.8 & 8.0    & 2.2    \\
   $P_{\mu}>0.2$~GeV    		&  40  &  39.3 	&  13.8   & 7.6  &  7.8 & 8.0    & 2.2    \\
   $N_{\mbox{decay-e}} <2$      &  34  &  36.1 	&  13.7   & 7.5  &  7.3 & 5.6    & 2.1    \\
	 Efficiency ($\%$)      	&      &  &  77.1   & 65.7 & 36.6 & 15.3   & 2.0     \\
   \hline
	 \hline
  \end{tabular}
\end{table}

{\it Oscillation fit.---}The oscillation parameters $\sin^2(\overline{\theta}_{23})$ 
and $\Delta\overline{m}^2_{32}$ are estimated using a maximum-likelihood fit to the measured
reconstructed energy spectrum in the far detector.
All other oscillation parameters are fixed as shown in Table~\ref{tab:oscpars}.
Oscillation probabilities are calculated using the full three-flavor oscillation framework~\cite{PhysRevD.22.2718}, assuming the normal mass hierarchy $(\Delta m^2_{32}>0)$.
Matter effects are included with an Earth density of $\rho$ = 2.6~g/cm$^3$~\cite{Hagiwara:2011}.

\begin{table}[!tbp]
    \centering
    \caption{Oscillation parameters used for the fit. 
The parameters $\sin^2(\overline{\theta}_{23})$ and $\Delta\overline{m}^2_{32}$ were allowed to fit in the ranges given. All other parameters were fixed to the values shown, taken from previous T2K fits~\cite{Abe:2015awa} and the Particle Data Group review~\cite{Agashe:2014kda}.}
    \label{tab:oscpars}
    \begin{tabular*}{0.9\columnwidth}{c@{\extracolsep{\fill}} c@{\extracolsep{\fill}} c}
        \hline
        \hline
        Parameter & $\nu$ & $\overline{\nu}$ \\
        \hline
        $\sin^2(\theta_{23})$ & 0.527 &  fit 0 -- 1  \\
        $\Delta m^2_{32}$ $(10^{-3}$~eV$^2)$ & 2.51 &  fit 0 -- 20 \\
        $\sin^2(\theta_{13})$ & \multicolumn{2}{c}{0.0248} \\
        $\sin^2(\theta_{12})$ & \multicolumn{2}{c}{0.304} \\
        $\Delta m^2_{21}$ $(10^{-5}$~eV$^2)$ & \multicolumn{2}{c}{7.53} \\
        $\delta_{CP}$ (rad) & \multicolumn{2}{c}{-1.55} \\
        \hline
        \hline
    \end{tabular*}
\end{table}

Confidence regions are constructed for the oscillation
parameters using the constant $\Delta \chi^2$ method~\cite{Agashe:2014kda}.
A marginal likelihood is used for this, integrating over the nuisance parameters $\mathbf{f}$ with prior probability functions $\pi(\mathbf{f})$ to find the likelihood as a function of only the relevant oscillation parameters $\mathbf{o}$:

\begin{equation}
\mathcal{L}(\mathbf{o}) = \int \prod_{i}^{\mathrm{E bins}} \mathcal{L}_{i}(\mathbf{o},\mathbf{f}) \times \pi(\mathbf{f}) \,d\mathbf{f},
\end{equation}
where Ebins denotes the number of reconstructed neutrino energy bins.

We define $\Delta \chi^2 = -2 \ln[\mathcal{L}(\mathbf{o})/\mathrm{max}(\mathcal{L})]$ 
as the ratio of the marginal likelihood at a point 
$\mathbf{o}$ in the $\sin^2(\overline{\theta}_{23})$ -- $\Delta\overline{m}^2_{32}$ oscillation
parameter space and the maximum marginal likelihood.
The confidence region is then defined as the area of the oscillation
parameter space for which $\Delta \chi^2$ is less than a standard critical value.
The Feldman-Cousins critical chi-square value was calculated for a coarse set of points in the oscillation parameter space.
The difference 
in the confidence region calculated from these points
and that from
the standard chi-square values was found to be negligible.

Table~\ref{tab:oa_syst} summarizes the fractional error on the expected number of SK events from a 1$\sigma$ variation of the flux, cross-section, and far-detector systematic parameters.
Although the fractional error on the expected number of events due to systematic errors is large, the effect of systematic parameters on the confidence regions found in this fit is negligible due to the limited data statistics.
The impact of fixing the values of
$\sin^{2}(\theta_{23})$ and $\Delta m^{2}_{32}$ in the fit is also negligible.

\begin{table}
\centering
\caption{Percentage change in the number of one-ring $\mu$-like events before the oscillation fit from 1$\sigma$ systematic parameter variations, assuming the oscillation parameters listed in Table~\ref{tab:oscpars} and that the antineutrino and neutrino oscillation parameters are identical.}
\label{tab:oa_syst}
\begin{tabular}{lc}
\hline
\hline
Source of uncertainty (number of parameters) & $\delta n^{\rm{exp}}_{\rm{SK}}$/$n^{\rm{exp}}_{\rm{SK}}$\\
\hline
ND280-unconstrained cross section (6) & 10.0\% \\
Flux and ND280-constrained cross section (31) & 3.4\% \\
Super-Kamiokande detector systematics (6) & 3.8\% \\
Pion FSI and reinteractions (6) & 2.1\% \\
\hline
Total (49) & 11.6\%\\
\hline
\hline
\end{tabular}
\end{table}

The observed $\overline{\nu}_\mu$ reconstructed energy spectrum from the antineutrino-beam-mode data is shown
in the upper plot of Fig.~\ref{fig:erec_ratio}, overlaid with the best-fit spectrum assuming normal hierarchy, separated by interaction mode.
The lower plot in Fig.~\ref{fig:erec_ratio} is the ratio of data to the expected, unoscillated spectrum.

The best fit values obtained
are $\sin^2(\overline{\theta}_{23}) = 0.45$ and 
$|\Delta\overline{m}_{32}^2| = 2.51\times10^{-3}{\mathrm{eV}^2}$, with 68\% confidence intervals of 0.38 -- 0.64 and 2.26 -- 2.80 $\times10^{-3}$~eV$^2$, respectively. 
A goodness-of-fit test was performed by comparing this fit to an ensemble of toy experiments, giving a $p$-value of 0.38.

\begin{figure}
\centering
\includegraphics[width=9cm]{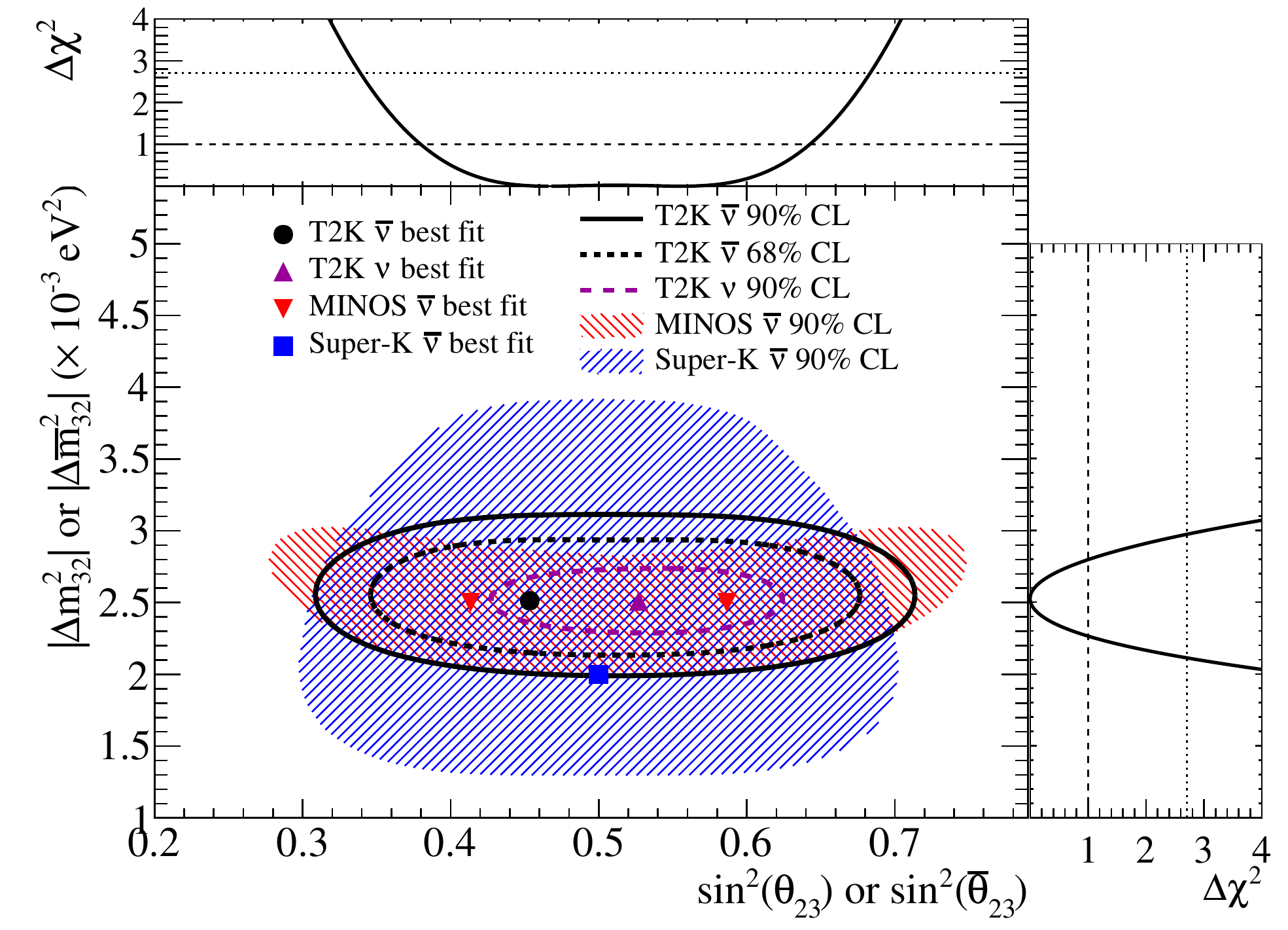}
\caption{The 68\% and 90\% confidence regions for $\sin^2(\overline{\theta}_{23})$ $-$ $|\Delta\overline{m}_{32}^2|$ plane assuming normal hierarchy, 
alongside 
the T2K $\nu$~\cite{Abe:2015awa}, SK Collaboration $\overline\nu$~\cite{Abe:2011ph}, and MINOS $\overline\nu$~\cite{Adamson:2012rm} 90\% confidence regions.
One-dimensional $\Delta\chi^2$ profiles for the two parameters 
are shown at the top and right,
overlaid with lines representing the 1D $\Delta\chi^2$ values 
for 68\% and 90\% confidence intervals.
}
\label{fig:contour}
\end{figure}

The fit results are shown in Fig.~\ref{fig:contour} as 68\% and 90\% confidence regions in the
$\sin^2(\overline{\theta}_{23})$ -- $\Delta\overline{m}^2_{32}$ plane.
The 90\% confidence regions from the T2K neutrino-beam-mode joint disappearance and appearance fit~\cite{Abe:2015awa},
the SK fit to $\overline\nu_\mu$ in atmospheric neutrino data~\cite{Abe:2011ph},
and the MINOS fit to $\overline{\nu}_\mu$ beam and atmospheric data~\cite{Adamson:2012rm} are also shown for comparison.
A second, fully Bayesian, analysis was also performed, 
producing a credible region matching the confidence regions presented above.

{\it Conclusions.---}We report the first study of $\overline{\nu}_{\mu}$ disappearance using an off-axis beam and present 
measurements of $\sin^{2}(\overline{\theta}_{23}) = 0.45$ and $\Delta \overline{m}^{2}_{32} = 2.51 \times 10^{-3}$ eV$^{2}$.
These results are consistent with the values of $\sin^{2}(\theta_{23})$ and $\Delta m^{2}_{32}$ observed previously by 
T2K~\cite{Abe:2015awa}, providing no indication of new physics, and are also in good agreement with similar measurements 
from MINOS~\cite{Adamson:2012rm} and the SK Collaboration~\cite{Abe:2011ph}.
The results presented here, with the first T2K antineutrino data set, are 
competitive with those from both MINOS and the SK Collaboration, 
demonstrating the effectiveness of the off-axis beam technique.
The data related to this measurement can be found in~\cite{Abe:datarelease}.

\begin{acknowledgments}
We thank the J-PARC staff for superb accelerator performance and the CERN NA61 
Collaboration for providing valuable particle production data.
We acknowledge the support of MEXT, Japan; 
NSERC (Grant No. SAPPJ-2014-00031), NRC and CFI, Canada;
CEA and CNRS/IN2P3, France;
DFG, Germany; 
INFN, Italy;
National Science Centre (NCN), Poland;
RSF, RFBR, and MES, Russia; 
MINECO and ERDF funds, Spain;
SNSF and SERI, Switzerland;
STFC, UK; and 
DOE, USA.
We also thank CERN for the UA1/NOMAD magnet, 
DESY for the HERA-B magnet mover system, 
NII for SINET4, 
the WestGrid and SciNet consortia in Compute Canada, 
and GridPP and the Emerald High Performance Computing facility
in the United Kingdom.
In addition, participation of individual researchers
and institutions has been further supported by funds from ERC (FP7), H2020 Grant No. RISE-GA644294-JENNIFER, EU; 
JSPS, Japan; 
Royal Society, UK; 
and the DOE Early Career program, USA.

%
\end{acknowledgments}

\bibliography{reference}

\providecommand{\noopsort}[1]{}\providecommand{\singleletter}[1]{#1}%
\begin{thebibliography}{34}%
\makeatletter
\providecommand \@ifxundefined [1]{%
 \@ifx{#1\undefined}
}%
\providecommand \@ifnum [1]{%
 \ifnum #1\expandafter \@firstoftwo
 \else \expandafter \@secondoftwo
 \fi
}%
\providecommand \@ifx [1]{%
 \ifx #1\expandafter \@firstoftwo
 \else \expandafter \@secondoftwo
 \fi
}%
\providecommand \natexlab [1]{#1}%
\providecommand \enquote  [1]{``#1''}%
\providecommand \bibnamefont  [1]{#1}%
\providecommand \bibfnamefont [1]{#1}%
\providecommand \citenamefont [1]{#1}%
\providecommand \href@noop [0]{\@secondoftwo}%
\providecommand \href [0]{\begingroup \@sanitize@url \@href}%
\providecommand \@href[1]{\@@startlink{#1}\@@href}%
\providecommand \@@href[1]{\endgroup#1\@@endlink}%
\providecommand \@sanitize@url [0]{\catcode `\\12\catcode `\$12\catcode
  `\&12\catcode `\#12\catcode `\^12\catcode `\_12\catcode `\%12\relax}%
\providecommand \@@startlink[1]{}%
\providecommand \@@endlink[0]{}%
\providecommand \url  [0]{\begingroup\@sanitize@url \@url }%
\providecommand \@url [1]{\endgroup\@href {#1}{\urlprefix }}%
\providecommand \urlprefix  [0]{URL }%
\providecommand \Eprint [0]{\href }%
\providecommand \doibase [0]{http://dx.doi.org/}%
\providecommand \selectlanguage [0]{\@gobble}%
\providecommand \bibinfo  [0]{\@secondoftwo}%
\providecommand \bibfield  [0]{\@secondoftwo}%
\providecommand \translation [1]{[#1]}%
\providecommand \BibitemOpen [0]{}%
\providecommand \bibitemStop [0]{}%
\providecommand \bibitemNoStop [0]{.\EOS\space}%
\providecommand \EOS [0]{\spacefactor3000\relax}%
\providecommand \BibitemShut  [1]{\csname bibitem#1\endcsname}%
\let\auto@bib@innerbib\@empty
\bibitem [{\citenamefont {Maki}\ \emph {et~al.}(1962)\citenamefont {Maki},
  \citenamefont {Nakagawa},\ and\ \citenamefont {Sakata}}]{Maki:1962mu}%
  \BibitemOpen
  \bibfield  {author} {\bibinfo {author} {\bibfnamefont {Z.}~\bibnamefont
  {Maki}}, \bibinfo {author} {\bibfnamefont {M.}~\bibnamefont {Nakagawa}}, \
  and\ \bibinfo {author} {\bibfnamefont {S.}~\bibnamefont {Sakata}},\ }\href
  {\doibase 10.1143/PTP.28.870} {\bibfield  {journal} {\bibinfo  {journal}
  {Prog. Theor. Phys.}\ }\textbf {\bibinfo {volume} {28}},\ \bibinfo {pages}
  {870} (\bibinfo {year} {1962})}\BibitemShut {NoStop}%
\bibitem [{\citenamefont {Pontecorvo}(1967)}]{Pontecorvo:1967fh}%
  \BibitemOpen
  \bibfield  {author} {\bibinfo {author} {\bibfnamefont {B.}~\bibnamefont
  {Pontecorvo}},\ }\href {http://www.jetp.ac.ru/cgi-bin/dn/e_026_05_0984.pdf}
  {\bibfield  {journal} {\bibinfo  {journal} {Zh. Eksp. Teor. Fiz.}\ }\textbf
  {\bibinfo {volume} {53}},\ \bibinfo {pages} {1717} (\bibinfo {year}
  {1967})},\ \bibinfo {note} {[Sov. Phys. JETP {\bf 26}, 984
  (1968)]}\BibitemShut {NoStop}%
\bibitem [{\citenamefont {Nakamura}\ \emph {et~al.}(2012)\citenamefont
  {Nakamura}, \citenamefont {Petcov} \emph
  {et~al.}}]{PhysRevD.86.010001neutrino}%
  \BibitemOpen
  \bibfield  {author} {\bibinfo {author} {\bibfnamefont {K.}~\bibnamefont
  {Nakamura}}, \bibinfo {author} {\bibfnamefont {S.~T.}\ \bibnamefont
  {Petcov}},  \emph {et~al.} (\bibinfo {collaboration} {Particle Data Group}),\
  }\href {\doibase 10.1103/PhysRevD.86.010001} {\bibfield  {journal} {\bibinfo
  {journal} {Phys. Rev. D}\ }\textbf {\bibinfo {volume} {86}},\ \bibinfo
  {pages} {010001} (\bibinfo {year} {2012})},\ \bibinfo {note} {see section 13.
  NEUTRINO MASS, MIXING, AND OSCILLATIONS.}\BibitemShut {Stop}%
\bibitem [{\citenamefont {Adamson}\ \emph {et~al.}(2012)\citenamefont {Adamson}
  \emph {et~al.}}]{Adamson:2012rm}%
  \BibitemOpen
  \bibfield  {author} {\bibinfo {author} {\bibfnamefont {P.}~\bibnamefont
  {Adamson}} \emph {et~al.} (\bibinfo {collaboration} {MINOS Collaboration}),\
  }\href {\doibase 10.1103/PhysRevLett.108.191801} {\bibfield  {journal}
  {\bibinfo  {journal} {Phys. Rev. Lett.}\ }\textbf {\bibinfo {volume} {108}},\
  \bibinfo {pages} {191801} (\bibinfo {year} {2012})}\BibitemShut {NoStop}%
\bibitem [{\citenamefont {Abe}\ \emph {et~al.}(2011{\natexlab{a}})\citenamefont
  {Abe} \emph {et~al.}}]{Abe:2011ph}%
  \BibitemOpen
  \bibfield  {author} {\bibinfo {author} {\bibfnamefont {K.}~\bibnamefont
  {Abe}} \emph {et~al.} (\bibinfo {collaboration} {Super-Kamiokande
  Collaboration}),\ }\href {\doibase 10.1103/PhysRevLett.107.241801} {\bibfield
   {journal} {\bibinfo  {journal} {Phys. Rev. Lett.}\ }\textbf {\bibinfo
  {volume} {107}},\ \bibinfo {pages} {241801} (\bibinfo {year}
  {2011}{\natexlab{a}})}\BibitemShut {NoStop}%
\bibitem [{\citenamefont {Abe}\ \emph {et~al.}(2011{\natexlab{b}})\citenamefont
  {Abe} \emph {et~al.}}]{Abe:2011ks}%
  \BibitemOpen
  \bibfield  {author} {\bibinfo {author} {\bibfnamefont {K.}~\bibnamefont
  {Abe}} \emph {et~al.} (\bibinfo {collaboration} {T2K Collaboration}),\ }\href
  {\doibase 10.1016/j.nima.2011.06.067} {\bibfield  {journal} {\bibinfo
  {journal} {Nucl. Instrum. Methods}\ }\textbf {\bibinfo {volume} {A659}},\
  \bibinfo {pages} {106} (\bibinfo {year} {2011}{\natexlab{b}})}\BibitemShut
  {NoStop}%
\bibitem [{\citenamefont {{D. Beavis, A. Carroll, I. Chiang, et
  al.}}(1995)}]{beavis:bnl}%
  \BibitemOpen
  \bibfield  {author} {\bibinfo {author} {\bibnamefont {{D. Beavis, A. Carroll,
  I. Chiang, et al.}}} (\bibinfo {collaboration} {E889 Collaboration}),\ }\href
  {http://www.hep.princeton.edu/~mcdonald/nufact/e889/authors.pdf} {\bibfield
  {journal} {\bibinfo  {journal} {{Physics Design Report}}\ }\textbf {\bibinfo
  {volume} {BNL 52459}} (\bibinfo {year} {1995})}\BibitemShut {NoStop}%
\bibitem [{\citenamefont {Abe}\ \emph {et~al.}(2012)\citenamefont {Abe} \emph
  {et~al.}}]{Abe2012}%
  \BibitemOpen
  \bibfield  {author} {\bibinfo {author} {\bibfnamefont {K.}~\bibnamefont
  {Abe}} \emph {et~al.} (\bibinfo {collaboration} {T2K Collaboration}),\ }\href
  {\doibase 10.1016/j.nima.2012.03.023} {\bibfield  {journal} {\bibinfo
  {journal} {Nucl. Instrum. Methods}\ }\textbf {\bibinfo {volume} {A694}},\
  \bibinfo {pages} {211} (\bibinfo {year} {2012})}\BibitemShut {NoStop}%
\bibitem [{\citenamefont {Abgrall}\ \emph {et~al.}(2011)\citenamefont {Abgrall}
  \emph {et~al.}}]{Abgrall:2010hi}%
  \BibitemOpen
  \bibfield  {author} {\bibinfo {author} {\bibfnamefont {N.}~\bibnamefont
  {Abgrall}} \emph {et~al.} (\bibinfo {collaboration} {T2K ND280 TPC
  Collaboration}),\ }\href {\doibase 10.1016/j.nima.2011.02.036} {\bibfield
  {journal} {\bibinfo  {journal} {Nucl. Instrum. Methods}\ }\textbf {\bibinfo
  {volume} {A637}},\ \bibinfo {pages} {25} (\bibinfo {year}
  {2011})}\BibitemShut {NoStop}%
\bibitem [{\citenamefont {Amaudruz}\ \emph {et~al.}(2012)\citenamefont
  {Amaudruz} \emph {et~al.}}]{Amaudruz:2012pe}%
  \BibitemOpen
  \bibfield  {author} {\bibinfo {author} {\bibfnamefont {P.}~\bibnamefont
  {Amaudruz}} \emph {et~al.} (\bibinfo {collaboration} {T2K ND280 FGD
  Collaboration}),\ }\href {\doibase 10.1016/j.nima.2012.08.020} {\bibfield
  {journal} {\bibinfo  {journal} {Nucl. Instrum. Methods}\ }\textbf {\bibinfo
  {volume} {A696}},\ \bibinfo {pages} {1} (\bibinfo {year} {2012})}\BibitemShut
  {NoStop}%
\bibitem [{\citenamefont {Fukuda}\ \emph {et~al.}(2003)\citenamefont {Fukuda}
  \emph {et~al.}}]{Fukuda2003418}%
  \BibitemOpen
  \bibfield  {author} {\bibinfo {author} {\bibfnamefont {S.}~\bibnamefont
  {Fukuda}} \emph {et~al.} (\bibinfo {collaboration} {Super-Kamiokande
  Collaboration}),\ }\href {\doibase 10.1016/S0168-9002(03)00425-X} {\bibfield
  {journal} {\bibinfo  {journal} {Nucl. Instrum. Methods}\ }\textbf {\bibinfo
  {volume} {A501}},\ \bibinfo {pages} {418} (\bibinfo {year}
  {2003})}\BibitemShut {NoStop}%
\bibitem [{\citenamefont {Abe}\ \emph {et~al.}(2014)\citenamefont {Abe} \emph
  {et~al.}}]{Abe20140211}%
  \BibitemOpen
  \bibfield  {author} {\bibinfo {author} {\bibfnamefont {K.}~\bibnamefont
  {Abe}} \emph {et~al.} (\bibinfo {collaboration} {Super-Kamiokande
  Collaboration}),\ }\href {\doibase 10.1016/j.nima.2013.11.081} {\bibfield
  {journal} {\bibinfo  {journal} {Nucl. Instrum. Methods}\ }\textbf {\bibinfo
  {volume} {A737}},\ \bibinfo {pages} {253} (\bibinfo {year}
  {2014})}\BibitemShut {NoStop}%
\bibitem [{\citenamefont {Abe}\ \emph {et~al.}(2015)\citenamefont {Abe} \emph
  {et~al.}}]{Abe:2015awa}%
  \BibitemOpen
  \bibfield  {author} {\bibinfo {author} {\bibfnamefont {K.}~\bibnamefont
  {Abe}} \emph {et~al.} (\bibinfo {collaboration} {T2K Collaboration}),\ }\href
  {\doibase 10.1103/PhysRevD.91.072010} {\bibfield  {journal} {\bibinfo
  {journal} {Phys. Rev.}\ }\textbf {\bibinfo {volume} {D91}},\ \bibinfo {pages}
  {072010} (\bibinfo {year} {2015})}\BibitemShut {NoStop}%
\bibitem [{\citenamefont {Abe}\ \emph {et~al.}(2013)\citenamefont {Abe} \emph
  {et~al.}}]{Abe:2012av}%
  \BibitemOpen
  \bibfield  {author} {\bibinfo {author} {\bibfnamefont {K.}~\bibnamefont
  {Abe}} \emph {et~al.} (\bibinfo {collaboration} {T2K Collaboration}),\ }\href
  {\doibase 10.1103/PhysRevD.87.012001} {\bibfield  {journal} {\bibinfo
  {journal} {Phys. Rev.}\ }\textbf {\bibinfo {volume} {D87}},\ \bibinfo {pages}
  {012001} (\bibinfo {year} {2013})},\ \bibinfo {note} {[Addendum: Phys.
  Rev.D87,no.1,019902(2013)]}\BibitemShut {NoStop}%
\bibitem [{\citenamefont {Ferrari}\ \emph {et~al.}(2005)\citenamefont
  {Ferrari}, \citenamefont {Sala}, \citenamefont {Fasso},\ and\ \citenamefont
  {Ranft}}]{cite:FLUKA1}%
  \BibitemOpen
  \bibfield  {author} {\bibinfo {author} {\bibfnamefont {A.}~\bibnamefont
  {Ferrari}}, \bibinfo {author} {\bibfnamefont {P.~R.}\ \bibnamefont {Sala}},
  \bibinfo {author} {\bibfnamefont {A.}~\bibnamefont {Fasso}}, \ and\ \bibinfo
  {author} {\bibfnamefont {J.}~\bibnamefont {Ranft}},\ }\href
  {http://www.osti.gov/scitech/biblio/877507} {\bibfield  {journal} {\bibinfo
  {journal} {{Report No. CERN-2005-010 and SLAC-R-773 and INFN-TC-05-11}}\ }
  (\bibinfo {year} {2005})}\BibitemShut {NoStop}%
\bibitem [{\citenamefont {B{\"o}hlen}\ \emph {et~al.}(2014)\citenamefont
  {B{\"o}hlen} \emph {et~al.}}]{FLUKA2011}%
  \BibitemOpen
  \bibfield  {author} {\bibinfo {author} {\bibfnamefont {T.~T.}\ \bibnamefont
  {B{\"o}hlen}} \emph {et~al.},\ }\href
  {http://ac.els-cdn.com/S0090375214005018/1-s2.0-S0090375214005018-main.pdf?_tid=8b9675ae-6fdb-11e5-ba8d-00000aab0f6b&acdnat=1444542623_86ffe2c6f1f08498777efcf7eb59937a}
  {\bibfield  {journal} {\bibinfo  {journal} {Nuclear Data Sheets}\ }\textbf
  {\bibinfo {volume} {120}},\ \bibinfo {pages} {211} (\bibinfo {year}
  {2014})}\BibitemShut {NoStop}%
\bibitem [{\citenamefont {Brun}\ \emph {et~al.}(1994)\citenamefont {Brun},
  \citenamefont {Carminati},\ and\ \citenamefont {Giani}}]{GEANT3}%
  \BibitemOpen
  \bibfield  {author} {\bibinfo {author} {\bibfnamefont {R.}~\bibnamefont
  {Brun}}, \bibinfo {author} {\bibfnamefont {F.}~\bibnamefont {Carminati}}, \
  and\ \bibinfo {author} {\bibfnamefont {S.}~\bibnamefont {Giani}},\ }\href
  {http://www.google.co.jp/url?sa=t&rct=j&q=&esrc=s&source=web&cd=1&ved=0CB0QFjAAahUKEwjV_sW33LnIAhXEnZQKHTAnCcU&url=http%3A%2F%2Fhepd.pnpi.spb.ru%2F~tatianas%2Fcernlib%2Fdoc%2Fgeant%2Faaaa.ps.gz&usg=AFQjCNFCH6HjSbTxhvm5_gfv3FguE1desA&sig2=kf0RuxPkmB-k9ZtIEH1Lgw}
  {\bibfield  {journal} {\bibinfo  {journal} {Report No. CERN-W5013}\ }
  (\bibinfo {year} {1994})}\BibitemShut {NoStop}%
\bibitem [{\citenamefont {Zeitnitz}\ and\ \citenamefont
  {Gabriel}()}]{cite:GCALOR}%
  \BibitemOpen
  \bibfield  {author} {\bibinfo {author} {\bibfnamefont {C.}~\bibnamefont
  {Zeitnitz}}\ and\ \bibinfo {author} {\bibfnamefont {T.~A.}\ \bibnamefont
  {Gabriel}},\ }\href@noop {} {\ }\bibinfo {note} {{\it In Proc. of
  International Conference on Calorimetry in High Energy Physics, Tallahassee,
  FL, USA, February 1993.}}\BibitemShut {Stop}%
\bibitem [{\citenamefont {Abgrall}\ \emph {et~al.}(2016)\citenamefont {Abgrall}
  \emph {et~al.}}]{Abgrall:2015xoa}%
  \BibitemOpen
  \bibfield  {author} {\bibinfo {author} {\bibfnamefont {N.}~\bibnamefont
  {Abgrall}} \emph {et~al.} (\bibinfo {collaboration} {NA61/SHINE
  Collaboration}),\ }\href {\doibase 10.1140/epjc/s10052-016-3898-y} {\bibfield
   {journal} {\bibinfo  {journal} {Eur. Phys. J.}\ }\textbf {\bibinfo {volume}
  {C76}},\ \bibinfo {pages} {84} (\bibinfo {year} {2016})}\BibitemShut
  {NoStop}%
\bibitem [{\citenamefont {Hayato}(2009)}]{Hayato:2009zzz}%
  \BibitemOpen
  \bibfield  {author} {\bibinfo {author} {\bibfnamefont {Y.}~\bibnamefont
  {Hayato}},\ }\href
  {http://www.actaphys.uj.edu.pl/fulltext?series=Reg&vol=40&page=2477}
  {\bibfield  {journal} {\bibinfo  {journal} {Acta Phys. Polon.}\ }\textbf
  {\bibinfo {volume} {B40}},\ \bibinfo {pages} {2477} (\bibinfo {year}
  {2009})}\BibitemShut {NoStop}%
\bibitem [{Hay()}]{Hayato:2009zzzz}%
  \BibitemOpen
  \href@noop {} {\ }\bibinfo {note} {Version 5.3.2 of NEUT library is used,
  which includes (i) the multinucleon ejection model of Nieves {\it et
  al.}~\cite{Nieves:2011pp} and (ii) nuclear long-range correlations for CCQE
  interactions, treated in the random phase
  approximation~\cite{Nieves:2004wx}.}\BibitemShut {Stop}%
\bibitem [{\citenamefont {Nieves}\ \emph {et~al.}(2011)\citenamefont {Nieves},
  \citenamefont {Ruiz~Simo},\ and\ \citenamefont
  {Vicente~Vacas}}]{Nieves:2011pp}%
  \BibitemOpen
  \bibfield  {author} {\bibinfo {author} {\bibfnamefont {J.}~\bibnamefont
  {Nieves}}, \bibinfo {author} {\bibfnamefont {I.}~\bibnamefont {Ruiz~Simo}}, \
  and\ \bibinfo {author} {\bibfnamefont {M.~J.}\ \bibnamefont
  {Vicente~Vacas}},\ }\href {\doibase 10.1103/PhysRevC.83.045501} {\bibfield
  {journal} {\bibinfo  {journal} {Phys.\ Rev.\ C}\ }\textbf {\bibinfo {volume}
  {83}},\ \bibinfo {pages} {045501} (\bibinfo {year} {2011})}\BibitemShut
  {NoStop}%
\bibitem [{\citenamefont {Nieves}\ \emph {et~al.}(2004)\citenamefont {Nieves},
  \citenamefont {Amaro},\ and\ \citenamefont {Valverde}}]{Nieves:2004wx}%
  \BibitemOpen
  \bibfield  {author} {\bibinfo {author} {\bibfnamefont {J.}~\bibnamefont
  {Nieves}}, \bibinfo {author} {\bibfnamefont {J.~E.}\ \bibnamefont {Amaro}}, \
  and\ \bibinfo {author} {\bibfnamefont {M.}~\bibnamefont {Valverde}},\ }\href
  {http://journals.aps.org/prc/abstract/10.1103/PhysRevC.70.055503} {\bibfield
  {journal} {\bibinfo  {journal} {Phys.\ Rev.\ C}\ }\textbf {\bibinfo {volume}
  {{\bf 70}}},\ \bibinfo {pages} {055503} (\bibinfo {year} {2004})},\ \bibinfo
  {note} {[Erratum-ibid.\ C {\bf 72} (2005) 019902]}\BibitemShut {NoStop}%
\bibitem [{\citenamefont {Wilkinson}()}]{cite:Wilkinson:2015}%
  \BibitemOpen
  \bibfield  {author} {\bibinfo {author} {\bibfnamefont {C.}~\bibnamefont
  {Wilkinson}},\ }\href
  {http://pos.sissa.it/archive/conferences/226/104/NUFACT2014_104.pdf}
  {\bibinfo  {journal} {{\it In Proc. of 16th International Workshop on
  Neutrino Factories and Future Neutrino Beam Facilities (NUFACT 2014),
  Glasgow, Scotland, UK, August 2014}}\ }\BibitemShut {NoStop}%
\bibitem [{\citenamefont {Llewellyn~Smith}(1972)}]{LlewellynSmith:1971zm}%
  \BibitemOpen
\bibfield  {journal} {  }\bibfield  {author} {\bibinfo {author} {\bibfnamefont
  {C.~H.}\ \bibnamefont {Llewellyn~Smith}},\ }\href {\doibase
  10.1016/0370-1573(72)90010-5} {\bibfield  {journal} {\bibinfo  {journal}
  {Phys. Rept.}\ }\textbf {\bibinfo {volume} {3}},\ \bibinfo {pages} {261}
  (\bibinfo {year} {1972})}\BibitemShut {NoStop}%
\bibitem [{\citenamefont {Jacob}(1978)}]{Jacob}%
  \BibitemOpen
  \bibfield  {author} {\bibinfo {author} {\bibfnamefont {M.}~\bibnamefont
  {Jacob}},\ }\href@noop {} {\emph {\bibinfo {title} {{Gauge Theories and
  Neutrino Physics}}}}\ (\bibinfo  {publisher} {{Elsevier Science Ltd}},\
  \bibinfo {address} {{North-holland/amsterdam}},\ \bibinfo {year}
  {{1978}})\BibitemShut {NoStop}%
\bibitem [{\citenamefont {Aguilar-Arevalo~{\it et al.}
  [MiniBooNE~Collaboration]}(2010)}]{AguilarArevalo:2010zc}%
  \BibitemOpen
  \bibfield  {author} {\bibinfo {author} {\bibfnamefont {A.~A.}\ \bibnamefont
  {Aguilar-Arevalo~{\it et al.} [MiniBooNE~Collaboration]}},\ }\href {\doibase
  10.1103/PhysRevD.81.092005} {\bibfield  {journal} {\bibinfo  {journal}
  {Phys.\ Rev.\ D}\ }\textbf {\bibinfo {volume} {{\bf 81}}},\ \bibinfo {pages}
  {092005} (\bibinfo {year} {2010})}\BibitemShut {NoStop}%
\bibitem [{\citenamefont {Aguilar-Arevalo}\ \emph {et~al.}(2013)\citenamefont
  {Aguilar-Arevalo} \emph {et~al.}}]{AguilarArevalo:2013hm}%
  \BibitemOpen
  \bibfield  {author} {\bibinfo {author} {\bibfnamefont {A.~A.}\ \bibnamefont
  {Aguilar-Arevalo}} \emph {et~al.} (\bibinfo {collaboration} {MiniBooNE}),\
  }\href {\doibase 10.1103/PhysRevD.88.032001} {\bibfield  {journal} {\bibinfo
  {journal} {Phys. Rev.}\ }\textbf {\bibinfo {volume} {D88}},\ \bibinfo {pages}
  {032001} (\bibinfo {year} {2013})}\BibitemShut {NoStop}%
\bibitem [{\citenamefont {Fiorentini}\ \emph {et~al.}(2013)\citenamefont
  {Fiorentini} \emph {et~al.}}]{Fiorentini:2013ezn}%
  \BibitemOpen
  \bibfield  {author} {\bibinfo {author} {\bibfnamefont {G.}~\bibnamefont
  {Fiorentini}} \emph {et~al.} (\bibinfo {collaboration} {MINERvA
  Collaboration}),\ }\href {\doibase 10.1103/PhysRevLett.111.022502} {\bibfield
   {journal} {\bibinfo  {journal} {Phys. Rev. Lett.}\ }\textbf {\bibinfo
  {volume} {111}},\ \bibinfo {pages} {022502} (\bibinfo {year}
  {2013})}\BibitemShut {NoStop}%
\bibitem [{\citenamefont {Fields}\ \emph {et~al.}(2013)\citenamefont {Fields}
  \emph {et~al.}}]{Fields:2013zhk}%
  \BibitemOpen
  \bibfield  {author} {\bibinfo {author} {\bibfnamefont {L.}~\bibnamefont
  {Fields}} \emph {et~al.} (\bibinfo {collaboration} {MINERvA Collaboration}),\
  }\href {\doibase 10.1103/PhysRevLett.111.022501} {\bibfield  {journal}
  {\bibinfo  {journal} {Phys. Rev. Lett.}\ }\textbf {\bibinfo {volume} {111}},\
  \bibinfo {pages} {022501} (\bibinfo {year} {2013})}\BibitemShut {NoStop}%
\bibitem [{\citenamefont {Barger}\ \emph {et~al.}(1980)\citenamefont {Barger},
  \citenamefont {Whisnant}, \citenamefont {Pakvasa},\ and\ \citenamefont
  {Phillips}}]{PhysRevD.22.2718}%
  \BibitemOpen
  \bibfield  {author} {\bibinfo {author} {\bibfnamefont {V.}~\bibnamefont
  {Barger}}, \bibinfo {author} {\bibfnamefont {K.}~\bibnamefont {Whisnant}},
  \bibinfo {author} {\bibfnamefont {S.}~\bibnamefont {Pakvasa}}, \ and\
  \bibinfo {author} {\bibfnamefont {R.~J.~N.}\ \bibnamefont {Phillips}},\
  }\href {\doibase 10.1103/PhysRevD.22.2718} {\bibfield  {journal} {\bibinfo
  {journal} {Phys. Rev. D}\ }\textbf {\bibinfo {volume} {22}},\ \bibinfo
  {pages} {2718} (\bibinfo {year} {1980})}\BibitemShut {NoStop}%
\bibitem [{\citenamefont {Hagiwara}\ \emph {et~al.}(2011)\citenamefont
  {Hagiwara}, \citenamefont {Okamura},\ and\ \citenamefont
  {Senda}}]{Hagiwara:2011}%
  \BibitemOpen
  \bibfield  {author} {\bibinfo {author} {\bibfnamefont {K.}~\bibnamefont
  {Hagiwara}}, \bibinfo {author} {\bibfnamefont {N.}~\bibnamefont {Okamura}}, \
  and\ \bibinfo {author} {\bibfnamefont {K.}~\bibnamefont {Senda}},\ }\href
  {http://dx.doi.org/10.1007/JHEP09%282011%29082} {\bibfield  {journal}
  {\bibinfo  {journal} {Journal of High Energy Physics}\ }\textbf {\bibinfo
  {volume} {2011}},\ \bibinfo {eid} {82} (\bibinfo {year} {2011})}\BibitemShut
  {NoStop}%
\bibitem [{\citenamefont {Olive}\ \emph {et~al.}(2014)\citenamefont {Olive}
  \emph {et~al.}}]{Agashe:2014kda}%
  \BibitemOpen
  \bibfield  {author} {\bibinfo {author} {\bibfnamefont {K.~A.}\ \bibnamefont
  {Olive}} \emph {et~al.} (\bibinfo {collaboration} {Particle Data Group}),\
  }\href {\doibase 10.1088/1674-1137/38/9/090001} {\bibfield  {journal}
  {\bibinfo  {journal} {Chin. Phys.}\ }\textbf {\bibinfo {volume} {C38}},\
  \bibinfo {pages} {090001} (\bibinfo {year} {2014})}\BibitemShut {NoStop}%
\bibitem [{\citenamefont {Abe}\ \emph {et~al.}()\citenamefont {Abe} \emph
  {et~al.}}]{Abe:datarelease}%
  \BibitemOpen
  \bibfield  {author} {\bibinfo {author} {\bibfnamefont {K.}~\bibnamefont
  {Abe}} \emph {et~al.},\ }\href@noop {} {\enquote {\bibinfo {title} {{T2K
  public data}},}\ }\bibinfo {howpublished}
  {\url{http://t2k-experiment.org/results/t2kdata-numubardis-2015}}\BibitemShut
  {NoStop}%
\end{thebibliography}%

\end{document}